\documentclass[epj]{svjour}
\usepackage{color}
\usepackage{bm}
\usepackage{amsfonts,amssymb}
\usepackage{amsmath}
\usepackage{enumerate}
\def\mg{\bm g}
\def\mn{\bm n}
\def\mk{\bm k}

\def\mv{\bm v}

\def\mx{\bm x}
\def\eps{\varepsilon}
\def\boldnabla{\bm \nabla }
\def\V{\mathcal{V}}
\def\dRM{\mathrm{d}}
\allowdisplaybreaks

\usepackage[pdftex]{graphicx}  %\usepackage{graphics}
\usepackage{subfig} %%%%%%%%%%%%%%%%%%%%%%%%%%%%%%%%%%%%%%%%
\begin{document}

%\linenumbers %%%%%%%%%%%%%%%%%%%%%%%%%%%%%%%%
%
\title{Scaling behavior in interacting systems: joint effect of anisotropy and compressibility}
\author{M. Hnati\v{c}\inst{1,}\inst{2} \and G. Kalagov\inst{1,}\inst{3}
\and T. Lu\v{c}ivjansk\'{y}\inst{1}
}                     % Do not remove
\offprints{}          % Insert a name or remove this line
\institute{Faculty of Science, \v{S}af\'arik University, Moyzesova 16, 040 01 Ko\v{s}ice,
Slovakia \and
Bogolyubov Laboratory of Theoretical Physics, Joint Institute
for Nuclear Research, 141980  Dubna, Russian Federation
\and
Department of Theoretical Physics, St.Petersburg State University, Ulyanovskaya 1, 
198504 St.Petersburg, Petrodvorets, Russian Federation
 }
\date{Received: date / Revised version: date}
\abstract{
Motivated by the ubiquity of turbulent flows in realistic conditions, effects
of turbulent advection on two
models of classical non-linear systems are investigated. In particular, 
 we analyze model A (according to the Hohenberg-Halperin classification \cite{hohenberg}) of a
  non-conserved order 
 parameter and a model of the direct bond percolation process.
  Having two paradigmatic 
  representatives of distinct stochastic dynamics, our aim is to elucidate to what extent
 velocity fluctuations affect their scaling behavior.
The main emphasis is put on an interplay between
anisotropy and compressibility of the velocity flow on their respective scaling regimes.
 Velocity fluctuations are generated by means of the Kraichnan rapid-change model, in which
  the anisotropy is due to a distinguished spatial
  direction $\mn$ and a correlator of the velocity field obeys the Gaussian distribution law with
 prescribed statistical properties.
  As the main theoretical
 tool, the field-theoretic perturbative renormalization group is adopted. Actual calculations are performed 
 in the leading (one-loop) approximation. Having 
 obtained infra-red stable asymptotic regimes, we have found four possible
 candidates for macroscopically observable behavior for each model. In contrast to the isotropic case, anisotropy
 brings about enhancement of non-linearities and non-trivial regimes are proved to be more stable.
 % end of PACS codes
} %end of abstract
\maketitle
%
%-----------------------------------------------------------------------------------------------------------
\section{\label{sec:intro}Introduction}
%-----------------------------------------------------------------------------------------------------------
 Critical behavior is present in many important physical phenomena. 
 A genuine interest in critical fluids stems from a fact that it represents a
 typical non-linear problem and
as such proves to be a challenging and enriching problem both from a theoretical and 
 an experimental point of view.
 The main research focus is devoted to an analysis of specific physical, thermodynamic and transport
properties. Majority of studies in the past employed a model of a pure fluid without any additional
interactions. However, later it became clear that hydrodynamic effects cannot be ignored due to 
 a strong sensitivity
of such systems to external disturbances  \cite{ivanov,sep}. These effects come into play
via  coupling between a critical fluid and an external environment in which fluid motion sets in.
A substantial increase of
compressibility, which is given by the derivative $\left(\partial \rho /\partial p\right)_T$ 
(where $\rho$ is the density of a fluid, $p$ is the pressure, and $T$ is the temperature 
in system), brings about so-called stratification effect: as density grows under the fluid's own 
weight,  a distinguished 
direction $\mn$ in the medium emerges \cite{Anisimov,comp1,comp3}.  Naturally, vector $\mn$ 
is parallel to the vector of gravity force $\mg$ and also to the density gradient, i.e. 
$ \mn \, || \, \boldnabla \rho $.  

  The environment can have an additional effect and may affect behavior profoundly \cite{Jan97}.
 In addition to external forces, such as gravitational, magnetic or electric field,
  there is another conceivable mechanism leading to a
 non-equilibrium state: turbulence caused by shaking, stirring, and flow generation that
 is abundantly observed in many natural phenomena \cite{davidson,frisch,monin}. 
 From a general standpoint, turbulence is a rule rather than exception \cite{davidson}
 and in order to have a complete physical picture turbulence should be incorporated
   in a theoretical description.
 
 Critical phenomena in chaotically stirred mediums  are abundant 
 in nature  as well. Of particular interest are fluids in turbulent motion, which manifest 
 universal behavior known as the Kolmogorov scaling in the case of fully developed turbulence
 \cite{davidson,frisch,monin}.
 The effect of stirring by a fully developed turbulent flow has been investigated in numerous 
 papers, see for
 instance \cite{frisch,monin,FGV01,turbo,HHL16} and references therein. Research activity
 regarding turbulence is enormous, but still many important problems remain unsolved.
 In theoretical physics there are
  two accepted approaches that allow us to include turbulence into consideration: the synthetic 
  velocity ensemble and the stochastic Navier-Stokes equation.
 A typical example of the former is embodied in 
 the rapid-change Kraichnan  model and its descendants \cite{FGV01,kr1,kr2,kr3,kr4}. An underlying
 idea is
 to generate the velocity 
 field $ {\mv}(\mx, t)$ by means of a statistical ensemble  with prescribed
 properties, which are chosen in a suitable way \cite{FGV01,turbo}. 
 Usually one assumes the
  Gaussian distribution law with zero mean and a correlation function in the form
 $\propto \delta(t-t') |\mx - \mx'|^{\delta}$.
 The latter (microscopic) approach to a description of velocity field has its own
 dynamics governed by the Navier-Stokes equation \cite{turbo,nav1,jurcisin2}
 augmented with a proper stochastic term, which mimics
  an input of energy into a system. 
  This approach is motivated by a fluctuation  theory of critical phenomena
  \cite{nav1,pata}.
  Though this approach is more satisfying from the physical point
 of view, in this work we consider a specific modification of the Kraichnan model. 
 At a first sight, the Kraichnan model seems to be
 oversimplified  compared to realistic realizations of the velocity flows \cite{davidson}. 
 Nevertheless, it
 captures relevant
 physical information about advection processes \cite{FGV01,turbo,Kra68} and at the
 same time
 it allows for exact solutions and thus provide us with a possibility for a mutual cross-check
 between different methods. An additional advantage consists in
 a relatively easy incorporation of other effects
 as anisotropy, compressibility, finite correlation time, helicity, and so on \cite{HHL16,Ant06}.
 
 In this paper we consider a compressible version of the Kraichnan model with a presence of anisotropy.
 Note that the turbulent compressibility has its own interest of study, which is motivated
 mainly by astrophysical applications 
  \cite{shore,Kim05,Sahra09,Galtier11,Banerjee13,Banerjee16,Hadid17}.
 An additional effect (anisotropy) can be justified by the following 
 heuristic observations. A typical terrestrial experiment necessarily occurs in
  the gravitational field, which can be locally considered as homogeneous. Gravity thus makes
 one spatial direction exceptional.
  A construction of a 
 realistic model of turbulence then requires an inclusion of the large-scale anisotropy induced 
 by the gravity vector
 $\mg$. According to the classical Kolmogorov-Obukhov theory of fully developed turbulence,
 anisotropy introduced at large scales by the forcing dies out when the energy is transferred
 down to 
 smaller scales owing to the cascade mechanism \cite{davidson,frisch,monin,turbo}. A number of works 
 confirms this pattern
 for even correlation functions, thus giving some quantitative support to the hypothesis of the
 restored local isotropy in the inertial-range turbulence. This should be valid for both
  velocity and passively advected fields as well.  More precisely, exponents
describing
the inertial-range scaling exhibit universality and hierarchy related to a degree of 
 anisotropy, and the leading
contribution to an even function is given by the exponent from the isotropic shell. 
Nevertheless, the
anisotropy survives in the inertial range and manifests itself in odd correlation 
functions \cite{n1,n2,n3}. 
Anisotropic turbulent systems with distinguished direction $\mn$ were first studied using 
the renormalization
group approach in \cite{n4}. A generalization to the case of anisotropic turbulence with a 
passive admixture 
was put forth in \cite{n5,n6,n7}.

 The presented problem is tackled by a versatile method of the renormalization group (RG).
In last three decades this method has grown
  into an indispensable
  tool for anyone whose aim is to determine 
  critical behavior in classical and quantum-many particle
 systems in a quantitatively reliable manner.
   Additionally, the RG  method provides us with a general conceptual framework 
   in which paramount concepts of scale invariance and universality can be justified.
   Depending on  certain crude properties of the system (dimensionality of space
   $d$, nature of order parameter and symmetry) a large-scale behavior of a
 system can be categorized
 into universality classes.
 Within a given class all pertained systems exhibit the same asymptotic behavior in 
 the macroscopic region, which corresponds to an infrared (IR) domain of the
 theory \cite{nav1,Zinn,Amit}. 
 The crucial idea of the Wilson  RG scheme goes as follows \cite{WilKog74,ZinnRG}:
  first, collective
 modes of a system are split into fast and
 slow degrees of freedom according to their momenta in the Fourier space. Then
  the fast (or short-range) modes are integrated out what effectively results into a
  construction of an IR effective 
 theory for the slow (or long-range) degrees of freedom \cite{wegner}. The action 
 functional of the theory, describing IR
 asymptotics,  can be expanded in terms of the order parameter and its derivatives. A
 widely accepted form of
 this functional is the Landau-Ginzburg-Wilson functional, and
 the resulting field theory is known as  the $O(N)$ $\varphi^4$-theory \cite{nav1,Zinn}, where
  $N$ is a number of components of the order parameter.

Effective field theories facilitate use of various perturbative methods for a controlled approximate
 calculation of critical exponents. One of them is based on straightforward calculations 
  in a form of perturbation series at fixed spatial dimensions  $d=3$ or $d=2$ \cite{parisi}. An alternative
  systematic scheme
is performed in a formal $(d_*-\varepsilon)$-dimension space, where $d_*$ is the upper critical 
dimension of the theory. 
 Then, the critical exponents have the form of $\varepsilon$-expansion and they have to be 
 resumed to obtain correct quantitative
results for realistic space dimensions \cite{Zinn,sok1,sok2,sok3}. Calculations
can be performed for the other
 universal quantities (asymptotic amplitude ratios or the equation of state) as well.
 In order to establish the structure of possible scaling
regimes and make a reliable prediction about obtained numerical values of critical exponents for 
real physical systems,
where $\varepsilon \sim \delta  \ge 1$, additional methods of the Borel summation 
 are required \cite{sok1,sok2,sok3}.

Fundamental difficulties with the RG approach are twofold. First, a multi-loop renormalization
group analysis of complex models is a very demanding task from a
technical point of view (see \cite{kompa} for recent progress in multi-loop
calculations of famous $\varphi^4$-theory). Majority of calculations in stochastic dynamics were restricted
only to a two-loop order. 
Broadly speaking \cite{aakv03} technical difficulties of  a two-loop calculation in stochastic
dynamics is as difficult as that of four-loop in critical statics. The main reason
lies in a more involved form of propagators and a proliferation of tensor
structures from interaction vertices.
 The only exceptions to the rule are model A \cite{AA84,ANS08} and
 the incompressible Kraichnan model \cite{AABKV01}.
Second, a calculation of expressions for universal quantities relies
on some additional physical considerations in order to make a proper extrapolation to physical
values \cite{nav1,Zinn}.
However, it is not at all clear whether this is a completely safe procedure. 
Hence, to make progress there have been many attempts 
 to go beyond the perturbation schemes. The non-perturbative (functional) renormalization
 group (NPRG) 
 is a method that does not rely on any small parameters in the 
 studied model \cite{ZinnRG,delamotte,polonyi}. The
NPRG was originally applied to equilibrium models of statistical physics  and a functional 
formulation
 allows a straightforward generalization to the case of non-equilibrium systems. 
 In particular, model A \cite{A}, the
 stochastic Navier-Stokes  equation \cite{NS}  and a passive field coupled to the Kraichnan 
 model \cite{kre}.
The RG method has been successfully applied on the analysis of the dynamic critical 
phenomena: critical
singularities of relaxation and correlation times, transport coefficients, etc. 
\cite{nav1,Zinn,tauber}. 
 The authors of
Ref. \cite{hohenberg} used this method to the time-dependent $O(1)$ Ginzburg-Landau models and 
showed that the RG 
method is consistent with the earlier mode-coupling theory and dynamic scaling. One of the 
representatives of 
the $O(1)$ symmetric $\varphi^4$-models is model A of the critical dynamics. The $O(1)$ 
class consists of systems with short-range forces and a scalar order parameter. This class
comprises the three-dimensional Ising model, liquid-gas critical point, binary fluid 
 mixtures, and uniaxial ferromagnets. 

Standard models of critical dynamics
(for instance models A, B, and their generalizations, in Hohenberg-Halperin classification \cite{hohenberg,nav1,tauber}) based
on the Ginzburg-Landau approach
do not cover all possible types of intriguing dynamic phenomena. 
The reason is that
 there exist non-equilibrium systems for which it is not even possible
 to postulate a static Hamiltonian \cite{nav1}. One of the crucial differences between
 critical dynamics and non-equilibrium models is related to the validity of the fluctuation-dissipation
 theorem. It holds only in the critical models. In theory this leads to restrictions
 regarding properties of the random stirring force \cite{nav1,kubo66}. Therefore, for
 non-equilibrium models a different route has to be undertaken, and  
  we have to start at a more fundamental level.
 For instance, one has to use a 
 master equation \cite{tauber,kampen}, and try to derive an effective coarse-grained model. 
 Being a formidable task
 there are only few systems for which such approach is reliable and mathematically
 well-founded. At the end one frequently  derives a 
  time-evolution (Liouvilean or quasi-Hamiltonian) operator
 \cite{Odor04,THL05}. This operator basically carries an information about rates between
 different states in the system.
 
  To conclude, fluctuation-dissipation theorem asserts  that the intensity of order
parameter fluctuations is known to scale in the same way as the susceptibility.
  Models of critical dynamics are in thermal equilibrium and the fluctuation-dissipation
 theorem has to be satisfied for them \cite{nav1,kubo66}. However, this is not the case for
 non-equilibrium systems \cite{nav1,tauber,Zia95}. In order to compare a role of 
 dynamics (equilibrium vs
  non-equilibrium) we take as an example
  a paradigmatic model known in as
 directed-bond percolation (DP) \cite{dp1,dp2,dp3,HHL08}. A general feature 
of DP is that an agent (a particle) can propagate from one site to another in a distinguished
direction: 
the direction in which agents are mainly spread out.  The DP may also serve for explaining various 
models of disease spreading, stochastic reaction-diffusion processes on a lattice or the wetting of
porous material \cite{dp2,ex2,ex3}. The most important aspect of DP is presence of a 
non-equilibrium second-order phase transition
between the absorbing and the active phase. The absorbing (inactive) phase corresponds to 
the case where
the medium does not contain agents, whereas in the active phase 
 the number of agents constantly fluctuates around a constant value. 
 Isotropic turbulent fluctuations in this model were considered in \cite{an3,tom1}. The effect of
strongly anisotropic turbulent motion modeled by the  Avellaneda-Majda ensemble on the critical
behavior was investigated in \cite{am} and in presence of additional long-range
force in \cite{Jan97,Jan99,Jan08}. In the latter additional L\'evy jumps were
allowed, which share some similarities with the turbulent advection (presence of long-range
correlations in space).

The main aim of the present paper is to analyze the effect of turbulent mixing on the
critical behavior of systems belonging to  model A and the DP model, respectively.
 These two models might be regarded as the simplest models
 of interacting quantities.
 In addition, we assume that velocity flow is compressible and anisotropic, and
 we would like to estimate mutual interplay between advection and self-interactions within
 a given model.
 
 This paper is organized as follows. In Sec.~\ref{sec:model}, we
briefly describe the investigated models. In Sec.~\ref{sec:rg}, we discuss the process of 
renormalization and present the RG functions calculated in the leading one-loop approximation. 
We classify possible IR regimes in Sec.~\ref{sec:scaling} and give their physical
interpretation in
 Sec.~\ref{sec:phys}. 

%================================================================================================
%------------------------------------------------------------------------------------------------------------------
\section{\label{sec:model} Models}
%------------------------------------------------------------------------------------------------------------------
Our main focus is on
 the large-scale behavior  of a scalar field $\varphi = \varphi(\mx, t)$  governed 
by a stochastic differential equation
\begin{equation}  
  \nabla_t \varphi = \lambda \, {\partial}^2 \varphi - \lambda \, 
  \frac{\delta {\V}(\varphi)}{\delta \varphi}\biggl|_{\varphi(\mx)\rightarrow \varphi(\mx,t)} + \eta, 
  \label{eq:EQ}
\end{equation}
where the velocity field $ {\mv}(\mx, t)$ appears
via the Lagrangian derivative $\nabla_t = \partial_t + (\mv\cdot\boldnabla)$. Here,
  $v_i$ is the $i$-component of the velocity field.  
 Further, $\lambda$  is a kinetic (diffusion) coefficient, $\partial_i = \partial/\partial x_i$ is partial
 derivative,  ${\partial}^2 = \partial_i \partial_i$ is the
 Laplace operator (summation over repeated index is implied and
henceforth will always be implicitly assumed), interaction functional (potential term) ${\V}$
is specified below for both model A and 
DP model, respectively, and $\delta/\delta\varphi$ is a variational derivative.
 Let us note that in this formulation a field $\varphi$ is regarded as a passive field, i.e.
 it does not affect dynamics of the velocity field $\mv$.
Also be aware of different interpretation and properties of the field $\varphi$ in models A and 
DP \cite{tauber}.
 
In this work, the following two local polynomial expressions for ${\V}(\varphi)$
\begin{equation}
  \label{eq:pot1}
  {\V}_A(\varphi) =\int \left[ \frac{\tau}{2} {\varphi}^{2}  + \frac{g_1}{4!}\,  
  {\varphi}^{4}\right] {\dRM}^d x, 
\end{equation}  
and
\begin{equation}
  \label{eq:pot2}
  {\V}_{DP}(\varphi) =\int \left[ \frac{\tau}{2} {\varphi}^{2}  + \frac{g_1}{3!}\, 
  {\varphi}^{3}\right] {\dRM}^d x
\end{equation}
are analyzed.

We have explicitly written a space dimension $d$, since
 in what follows we employ dimensional regularization technique in which
  $d$ is considered to be a continuous variable \cite{nav1,Zinn,Amit}.
The suitable random field $\eta=\eta(\mx, t)$  in Eq.~(\ref{eq:EQ}) for each potential
(\ref{eq:pot1}) and (\ref{eq:pot2}) is chosen to obey the Gaussian distribution law. 
 Though studied models share some similarities there is a profound difference
  regarding their noise properties. 
The noise $\eta_{\,A}$ with zero mean  
 is fully specified by its second moment (correlation function)
\begin{equation}
  \label{eq:eta4}
  \langle\eta_{\,A}(\mx, t)\,\eta_{\,A}(\mx', t')\rangle = 2 \lambda 
  \, \delta(t-t')\,\delta(\mx - \mx'),
\end{equation}
and it is an example of additive noise \cite{kampen,gardiner}. A precise form follows from the
requirement that in the limit $t \rightarrow +\infty$ the system 
 has to  relax into a thermal equilibrium steady state \cite{nav1,kubo66}. 
 Statistical averaging $\langle\ldots\rangle$ in (\ref{eq:eta4}) runs over all possible
 realizations of the random field $\eta_A$ with suitable boundary conditions imposed \cite{nav1}.

 On the other hand, the DP model differs in presence of an absorbing state from which
 the system cannot escape. It is intuitively clear that this property has to be somehow
 carried to the coarse-grained description (\ref{eq:EQ}).
 It turns out  \cite{dp3}  that it leads
 to a multiplicative noise with a
 correlator $\eta_{\,DP}$  in the following form
\begin{equation}
  \label{eq:eta3}
  \langle\eta_{\,DP}(\mx, t)\,\eta_{\,DP}(\mx', t')\rangle = g_1 \,\lambda 
  \, \delta(t-t')\,\delta(\mx - \mx')\, \varphi(\mx, t).
\end{equation}
This correlator clearly
  reflects the absorbing state condition: the noise fluctuations cease
  in the absorbing state $\varphi = 0$ and
  it can be shown that the fluctuation-dissipation connection is lost \cite{dp3,HHL08}.

%%%%%%%%%%%%%%%%%%%%%%%%%%%%%%%%%%%%%%%%%%%%%%%%%%%%%%%%%%%%%%%%%%%%%%%%%%%%%%%%%%
%
%               ADDED IN REVISION
%
%%%%%%%%%%%%%%%%%%%%%%%%%%%%%%%%%%%%%%%%%%%%%%%%%%%%%%%%%%%%%%%%%%%%%%%%%%%%%%%%%%
{
From the fundamental point of view, the correlator (\ref{eq:eta3}) is not very persuasive.
 The left-hand side of the equation clearly corresponds to a statistical average, whereas the right-hand side is a random quantity.
 It would be possible to reformulate the percolation process in terms of multiplicative noise
 \cite{an1}. However, with regard to universal properties the ensuing field-theoretic action would be the same \cite{dp3}.
  In what follows, we thus remain in consensus with the existing literature and consider the correlator as
  a starting point for the following field-theoretic analysis.
}
%%%%%%%%%%%%%%%%%%%%%%%%%%%%%%%%%%%%%%%%%%%%%%%%%%%%%%%%%%%%%%%%%%%%%%%%%%%%%%%%%%
%
%               END OF REVISION
%
%%%%%%%%%%%%%%%%%%%%%%%%%%%%%%%%%%%%%%%%%%%%%%%%%%%%%%%%%%%%%%%%%%%%%%%%%%%%%%%%%%  
  
The couple $\{\eta_{\,A},\, {\mathcal V}_A(\varphi)\}$ fully specifies model A 
describing the critical dynamics of a non-conserved order parameter near the
 thermal equilibrium with respect to universal properties \cite{nav1,Zinn}. 
In a vicinity of critical temperature $T_c$ a control parameter
is usually chosen as a deviation $\tau \sim T - T_c $ from the corresponding critical value
  (see \cite{hohenberg,tauber,folk}). 
  
On the other hand, the couple $\{\eta_{\,DP},\, {\mathcal V}_{DP}(\varphi)\}$
fully incorporates universal properties of 
 a  non-equilibrium phase transition in the DP process. In this case 
$\tau \sim p - p_c$ is a deviation from percolation threshold probability $p_c$. 
 At $p = p_c$ the system exhibits the directed percolation phase transition \cite{tauber,dp2}
 from an active to an absorbing phase. So there is a formal analogy regarding $\tau$
 in these two cases.

 To finalize a theoretical description we have yet to prescribed properties of the velocity field $\mv$.
 As has been already mentioned in Sec.~\ref{sec:intro}
  we assume that the  velocity field obeys a Gaussian distribution.
  A non-zero mean value of velocity can always be subtracted via  a suitable redefinition
  of advected field $\varphi$. From now on we therefore assume that $\langle \mv \rangle = 0$.
  Since we are dealing with a translationally
 invariant theory it is convenient to provide the correlator directly in the Fourier (momentum) 
 representation
\begin{equation}
  \label{eq:corr}
  {\mathcal D}_{i\,j}({\mk}, \omega) \equiv \langle v_{i}({\mk}, \omega)
  \upsilon_{j}(-{\mk}, -\omega) \rangle =  \frac{ D_0 \, {\mathcal T}_{i\,j}({\mk})}{\left({k}^2
  + {\ell}^{-2}\right)^{(d + \delta)/2}}.
\end{equation}
Here,  $D_0 > 0$ is a positive amplitude, $k = |\mk|$ is a magnitude of a
 momentum vector $\mk$,  
 parameter $\ell$ is an external macroscopic  scale ($\ell$ has the
 same order of magnitude as a linear size of a system considered), which
 provides IR regularization of the theory $k \ell \gg 1$. Explicit calculations
 show independence of universal critical exponents on
  this scale $l$. An additional 
 exponent $\delta$ is a perturbation parameter that controls a deviation 
 from the Kolmogorov  turbulent regime, which corresponds to the value 
 $\delta_K = 4/3$ \cite{FGV01}. In the RG method it plays a role of an analytic
 regulator \cite{nav1}.
 If the considered system is isotropic 
 and incompressible, the tensor ${\mathcal T}_{i\,j}( \mk )$  
  equals to a transverse projection operator $\mathcal{P}\!_{i\,j}(\mk) 
 \equiv \delta_{i\,j} - k_{i} k_{j}/{ k}^2$.  This is in accordance with 
 the condition $\boldnabla\cdot\mv  = 0 $ for a  divergence-free velocity 
 field.   Let us consider a different case, for which there is a distinguished 
 spatial direction denoted 
 by a unit vector  $\mn$. Then the system possesses an uniaxial anisotropy,
 and the representation of ${\mathcal T}_{i\,j}(\mk)$ takes the following  tensor structure  
 \cite{n5}
\begin{align}
  {\mathcal T}_{i\,j}( \mk ) & =  \tilde{a}( \theta ) \mathcal{P}_{i\,j}( \mk ) 
  + \tilde{b} ( \theta ) \mathcal{P}\!_{i\, s}( \mk )n_{s} \, \mathcal{P}\!_{j\,l}( \mk )n_{l} 
  \nonumber\\
  & +\alpha \tilde{c}( \theta ) \mathcal{Q}_{i\,j}( \mk ),
\end{align}
where in addition the longitudinal tensor $\mathcal{Q}_{i\,j}( \mk ) \equiv  k_{i} k_{j}/{k}^2$
appears due to a conceivable compressible modes of the velocity 
 field ${\mv}$. Further, the angle variable $\theta$ denotes an angle
between the direction $ \mn $ and the wave vector $ \mk $ ($\mn\cdot\mk = k\cos\theta$).
Dimensionless scalar
functions $ \tilde{a}(\theta), \, \tilde{b}(\theta), \, \tilde{c}(\theta)$ can be decomposed into the
Gegenbauer polynomials, $d$-di\-men\-sio\-nal  generalization of the Legendre polynomials. To be 
more specific, in the following analysis we confine ourselves to the special choice 
(see \cite{n5}) 
\begin{equation}  
  \tilde{a}(\theta) = 1 + \rho_1 \cos^2(\theta), \quad \tilde{b}(\theta) = \rho_2,
 \quad \tilde{c}(\theta) = 1 + \rho_3 \cos^2(\theta).
\end{equation}  
  Anisotropy parameters $\rho_1, \rho_2, \rho_3$ are subject
to the inequalities
\begin{equation}
  \rho_1 > -1,\quad \rho_2 > -1,\quad \rho_3 > -1, 
\end{equation}
that ensures positive definiteness of the corresponding Gaussian kernel 
 (see Eq.~(\ref{eq:SV}) in the following). It has been
established \cite{n5} that
this case displays main features of anisotropy fluctuations. An arbitrary dimensionless
parameter $\alpha$ can attain any positive value. 

 The structure of the velocity field permits us to investigate the limit of a
 potential  flow that fulfills the condition  $ \boldnabla \times \mv = 0$. 
 Introducing a new finite parameter  $ \widetilde{D}_0 = D_0 \,\alpha$ and passing 
 to the limit $\alpha \to \infty $, we get
 the  longitudinal correlator for an irrotational velocity  flow in a form
 \begin{equation}
 \label{eq:rot}
 \quad  {\mathcal D}_{i\,j}(\mk, \omega) =  \frac{ \widetilde{D}_0 \, \tilde{c} ( \theta )
 \, {\mathcal Q}_{i\,j}(\mk)}{\left({k}^2 + {\ell}^{-2}\right)^{(d + \delta)/2}}.
\end{equation}
%----------------------------------------------------------------------------------------------------------
\section{\label{sec:rg} Renormalization procedure}
%----------------------------------------------------------------------------------------------------------
According to the general formalism \cite{nav1,tauber}, stochastic problems 
 (\ref{eq:EQ}--\ref{eq:eta3}) 
are tantamount to the  field models of the doubled set of fields $\{\varphi, \varphi'\}$. 
Main benefits of such reformulation are a transparent perturbation expansion and 
 an effective use of the RG method, which corresponds to an infinite resummation of given
classes of Feynman diagrams. That makes the RG method an especially powerful and versatile theoretical tool.

Effective theories for model (\ref{eq:EQ}--\ref{eq:eta3}) can be constructed in a
straightforward fashion. Field-theoretic
action functional ${\mathcal S}_{A}$ for model A reads
\begin{align}  
  {\mathcal S}_{A} & =  \varphi' \left[ - \nabla_t-  a  \left({\partial_i  \upsilon_i}\right) + 
  \lambda \, {\partial}^2 - \lambda \, \tau + \lambda \, b \,  {(n_i  \partial_i)}^2\right] 
  \varphi \nonumber \\
  & - \frac{g_1 \lambda }{3!} \varphi^3 \varphi' +  \lambda \varphi' \varphi' + {\mathcal S}({\mv}),
  \label{eq:SA} 
\end{align}
and ${\mathcal S}_{DP}$ for the model takes the following form
\begin{align}
  {\mathcal S}_{DP} & = \varphi' \left[ - \nabla_t-  a \left({\partial_i  \upsilon_i}\right) +
  \lambda \, {\partial}^2 - \lambda \, \tau  + \lambda\,b {(n_i \partial_i)}^2\right] \varphi 
  \nonumber\\ 
  & + \frac{g_1 \lambda}{2} ( \varphi' -  \varphi)\, \varphi' \varphi + {\mathcal S}({\mv}).
  \label{eq:SDP} 
\end{align}    
For brevity we have introduced a condensed notation, in which
integrals over whole space-time are implicitly included, e.g., the third
term in Eq.~(\ref{eq:SA}) on the right-hand side is an abbreviated form of the expression
\begin{equation}
   \varphi' \partial^2 \varphi = 
   \int\! \dRM t \!\int\! \dRM^d x \mbox{ }\varphi '(\mx,t) \boldnabla^2 \varphi(\mx,t). 
\end{equation}

Let us make a comment regarding terms $v_i\partial_i$ and $a(\partial_i v_i)$ appearing
in Eqs.~(\ref{eq:SA}) and (\ref{eq:SDP}).
 In the incompressible case the latter term is not present, whereas in the compressible case
 \cite{Ant06,Ant00} there are 
 two physically relevant and distinguishable cases
 \begin{enumerate}[a)] 
   \item passive advection of density field with a convective part in the form
	 \begin{equation}
	    \partial_t \varphi + \boldnabla\cdot (\mv\varphi) = \ldots,
	 \end{equation}
   \item passive advection of tracer field obtained as follows
	 \begin{equation}
	    \partial_t \varphi + (\mv\cdot\boldnabla) \varphi = \ldots,
	 \end{equation} 	 
 \end{enumerate}
where $ \dots $ stands for neglected diffusion and source terms (their presence is not relevant). 
 For a non-interacting scalar field $\varphi$ a distinction
 between cases a) and b)
 is preserved during perturbation procedure. As it has been pointed out
in \cite{an7} for interacting theories the situation differs on a fundamental level.
 In fact, both terms $\boldnabla\cdot (\mv\varphi)$ and
 $(\mv\cdot\boldnabla) \varphi$
 have to be included. Even if we did not include them, RG procedure would generate it. 
 Hence, in order to get a multiplicatively renormalizable model \cite{nav1,Zinn} we are forced to
 introduce such terms in our action functionals from the very beginning.
 Same reasoning leads to a presence of a term ${(n_i \partial_i)}^2 \varphi$ in actions 
 (\ref{eq:SA}) and (\ref{eq:SDP}). Indeed, it is easy to verify that 
 corresponding counterterms appear in the renormalization process starting with the lowest 
order of a perturbation expansion. From the formal point of view, we can interpret models (\ref{eq:SA}) and (\ref{eq:SDP}) as two-scale anisotropic models (one
scale connected with the time variable and the other with special spatial direction). This makes
an overall analysis even more complicated and cumbersome than what one expects
in typical dynamical model \cite{nav1,tauber}.

The free field-theoretic action for  the ${\mv}$-field reads \cite{turbo,kr4}
\begin{equation}
  \label{eq:SV}
  {\mathcal S}({\mv}) = -\frac{1}{2} \upsilon_i {\mathcal D}^{-1}_{\,i\,j} \upsilon_j,
\end{equation}
and is due to the  assumed Gaussian nature of velocity field $\mv$.
 A dimension analysis reveals that the $g_1$ vertices in the models are marginal at $d=4$. At 
the same time, an inclusion of the velocity field leads to a new non-linear term 
$\varphi' ({\upsilon_i  \partial_i}) \varphi$. This term generates vertices in perturbation expansion 
 containing momentum of the field  $\varphi$.
 
 %%%%%%%%%%%%%%%%%%%%%%%%%%%%%%%%%%%%%%%%%%%%%%%%%%%%%%%%%%%%%%%%%%%%%%%%%%%%%%%%%%
%
%               CHANGED IN REVISION
%
%%%%%%%%%%%%%%%%%%%%%%%%%%%%%%%%%%%%%%%%%%%%%%%%%%%%%%%%%%%%%%%%%%%%%%%%%%%%%%%%%%
 %Note that
 %in order to analyze
%the effect of the velocity field on thermal fluctuations in a critical domain, it is necessary
%to consider parameters $\varepsilon \equiv  4 - d$ and $\delta$ to be small and of the same 
%magnitude $\varepsilon \sim \delta$. Otherwise one of the vertices is IR
%irrelevant and should be discarded \cite{nav1} at the outset.
{
Note that in contrast to standard RG approaches, in the paper we deal with the two-parameter
expansion $(\varepsilon,\delta)$, where $\varepsilon \equiv  4 - d$ denotes a deviation from the 
 upper critical space dimension.
 Logarithmic theory is obtained
 for $\varepsilon = \delta = 0$ and UV (ultraviolet) divergences manifest themselves as poles in linear combinations $a\varepsilon+b\delta,$
  with $a,b$ being constants.
  }
%%%%%%%%%%%%%%%%%%%%%%%%%%%%%%%%%%%%%%%%%%%%%%%%%%%%%%%%%%%%%%%%%%%%%%%%%%%%%%%%%%
%
%               END OF REVISION
%
%%%%%%%%%%%%%%%%%%%%%%%%%%%%%%%%%%%%%%%%%%%%%%%%%%%%%%%%%%%%%%%%%%%%%%%%%%%%%%%%%%

 %%%%%%%%%%%%%%%%%%%%%%%%%%%%%%%%%%%%%%%%%%%%%%%%%%%%%%%%%%%%%%%%%%%%%%%%%%%%%%%%%%
%
%               ADDED IN REVISION
%
%%%%%%%%%%%%%%%%%%%%%%%%%%%%%%%%%%%%%%%%%%%%%%%%%%%%%%%%%%%%%%%%%%%%%%%%%%%%%%%%%%
{
Throughout this paper, actions (\ref{eq:SA}) and (\ref{eq:SDP}) should be interpreted in It\^o sense, i.e.,
 Heaviside step function $\theta(t)$ in propagators is set to zero for $t = 0$ \cite{nav1,dp3}.
 }
%%%%%%%%%%%%%%%%%%%%%%%%%%%%%%%%%%%%%%%%%%%%%%%%%%%%%%%%%%%%%%%%%%%%%%%%%%%%%%%%%%
%
%               END OF REVISION
%
%%%%%%%%%%%%%%%%%%%%%%%%%%%%%%%%%%%%%%%%%%%%%%%%%%%%%%%%%%%%%%%%%%%%%%%%%%%%%%%%%%

 A calculation of the RG functions proceeds in a standard fashion. As the main points 
are well-known \cite{nav1,Zinn,Amit} we refrain here from mentioning
all the intermediate steps. Let us thus proceed directly to beta functions, which
are fundamental quantities for determination of scale behavior.

 Elimination of all UV-divergences from one-loop 1-particle irreducible graphs 
 makes the theory UV finite. An important byproduct is a relation between UV and IR behavior
 in the renormalized theory \cite{nav1,tauber}. This
  yields a non-trivial information about scaling behavior
 via knowledge of flow beta-functions ($\beta$-functions) of the model.
 For technical reasons we have used
  the framework of the dimensional regularization in the 
  $\mathbb{R}^{4-\varepsilon} \times \mathbb{R}$-space 
(for more details see \cite{HHL16,nav1}). Let us proceed to explicit results. In order
to simplify the
notation, let us rescale the coupling constants according to
$ g_1/(16 \pi^2) \rightarrow g_1, \, g_2/(16 \pi^2) \rightarrow g_2$ for model A. We can then
 write the corresponding $\beta$-functions in the following way
\begin{align}
  \label{eq:betaA1}
  \beta_{a} & =   \frac{(4 a - 1 )}{4 \sqrt{b+1}}\, g_1,\\
  \label{eq:betaA2}
  \beta_{b} & =  -\frac{g_2}{12} (  \alpha\,  \rho_3 -  \rho_1 +  7 \rho_2) \nonumber \\  
  & + \frac{g_2 \, b }{24}
  \left[  \alpha\, (6 + \rho_3)+5 \rho_1+\rho_2+18 \right], \\  
  \beta_{g_1} &  =  - g_1 \, \varepsilon  +  \frac{3\,  g_1^2 }{ \sqrt{b+1}}+ g_1 g_2\biggl\{
   \frac{ \alpha  (6 + \rho_3)\! +\! 5 \rho_1 \! + \! \rho_2\! + \! 18 }{12}
  \nonumber \\ 
  &- \frac{\alpha (4 a^2-2 a+1 )}{b^2} \left[ 2 (\sqrt{b+1}-1) (b - \rho_3) + b \rho_3 \right] \biggl\},
  \label{eq:betaA3}
  \\
  \beta_{g_2} & = - \ g_2 \, \delta  + \frac{g_2^2 }{24} \left[  \alpha\, (6 + \rho_3)+5 \rho_1+\rho_2+18 \right].
  \label{eq:betaA4}
\end{align}
On the other hand,
 for the DP model it is more suitable to employ the
 following rescaling  
\begin{equation} 
  \frac{g^2_1}{16 \pi^2} \rightarrow g_1, \quad
  \frac{g_2}{16 \pi^2}
  \rightarrow g_2.
\end{equation}
Then the $\beta$-functions for DP model can be written as follows
\begin{align}
  \beta_{a} & =  \frac{(2 a - 1 )}{8 \sqrt{b+1}} \, g_1,
  \label{eq:betaG1}
  \\
  \beta_{b} & =  -\frac{g_2}{12} ( \alpha \rho_3 - \rho_1 + 7 \rho_2)+\frac{g_2\, b }{24} 
  \biggl[  \alpha (6 + \rho_3)+5 \rho_1 \nonumber\\
  & +\rho_2+18 \biggl],
  \label{eq:betaG2}
  \\
  \beta_{g_1} & = - g_1 \,\varepsilon  + \frac{g_1^2}{ 2 \sqrt{b+1}} + g_1g_2 \biggl\{ 
  \frac{  \alpha \, (6 + \rho_3)\! + \! 5 \rho_1\! +\! \rho_2\! +\! 18 }{12} \nonumber \\
  & - \frac{\alpha (2 a^2-2 a+1)}{b^2} \left[ 2 (\sqrt{b+1}-1) (b - \rho_3) + b \rho_3 \right]  
  \biggl\}
  , 
  \label{eq:betaG3}
  \\
  \beta_{g_2} & = - \ g_2 \, \delta  + \frac{g_2^2}{24} \left[  \alpha (6 + \rho_3)+5 
  \rho_1+\rho_2+18 \right] + \frac{g_1\, g_2}{8 \sqrt{b+1}}.  
  \label{eq:betaG4}
\end{align}
We observe that there is a certain similarity in above two sets of $\beta$-functions. However,
 distinct non-linearities lead to terms, which clearly differ from each other.
    
    Let us make a general remark. The parameters $a, b, \alpha$ 
    are not perturbation charges, although they appear in perturbative expansions.
    They present non-perturbative characteristics of the model whose
    values are not restricted (apart from additional
    physical considerations). Further, because of the symmetry of  model DP under transformations
 \begin{align}
  \varphi( \mx,t )  \rightarrow -\varphi'( \mx,-t ), \quad g_1 \rightarrow - g_1, \\  
   \varphi'( \mx,t )  \rightarrow -\varphi( \mx,-t ),  \quad  a  \rightarrow  1 - a,
\end{align}
we see that equality $\beta_{a} = 0$ holds for $a = 1/2$, and is valid to 
 all orders in the perturbation theory \cite{an7}.

%------------------------------------------------------------------------------------------------------------
{\section{ Critical scaling} \label{sec:scaling}}
%------------------------------------------------------------------------------------------------------------
The $\beta$-functions of the theory describe the evolution of the effective coupling
constants upon changing a  wave number scale. The first step in an analysis of the asymptotic
behavior of the models consists in a calculation of fixed points (FPs), whose
coordinates we briefly denote as $g^*\equiv(g_1^*, g_2^*, a^*, b^*)$.
 They are defined as a solution to the following set of interconnected equations
\begin{align}
  %\beta_{g_1}(g_1^*, g_2^*, a^*, b^*) &= 0,  &\beta_a&(g_1^*, g_2^*, a^*, b^*) =0,  \\
  %\beta_{g_2}(g_1^*, g_2^*, a^*, b^*) &= 0,  &\beta_b&(g_1^*, g_2^*, a^*, b^*) =0.
  \beta_{g_1}(g^*) =  \beta_a (g^*) =
  \beta_{g_2}(g^*) = \beta_b (g^*) =0.
\end{align}
In statistical physics we are especially interested in IR stable FP, i.e., in such a point for which
 matrix  $\Omega_{g g'} \equiv \partial \beta_{g}/\partial {g'}|_{g=g^{*}}$,
where $g, g' \in\{g_1, g_2, a, b\}$,    is a positive definite matrix \cite{nav1}. 
Such points are obvious candidates for macroscopically observable regimes and in
 principle can be experimentally realized \cite{nav1,tauber}.

The following section~\ref{sec:model_A} is devoted to fixed points' analysis of model A, whereas
in section~\ref{sec:model_DP} we discuss model DP.
%------------------------------------------------------------------------------------------------------------
{\subsection{Model A} \label{sec:model_A}}
%------------------------------------------------------------------------------------------------------------
	Let us analyze permissible fixed points %($g_1^*, \,g_{2}^*,\, a^*,\, b^*$) 
	($g^*$) of the 
	system (\ref{eq:betaA1})-(\ref{eq:betaA4}) at physical values of perturbation parameters 
	$\varepsilon = 1$, and $\delta_K = 4/3$. In order to determine a type of a fixed point, 
	eigenvalues $\Omega_l, l=1,\dots,4$ of the matrix $\Omega$ have to be analyzed. 
	In what follows we use the following notation 
	$\Omega = \left[\Omega_1, \,\Omega_2,\, \Omega_3, \,\Omega_4  \right]$ for 
	four eigenvalues at a given fixed point. In general, we deal with non-diagonal matrices, but
	in all calculations it was possible to compute eigenvalues in an explicit form using
	symbolic software packages \cite{maple}.
        
        Altogether we have found four fixed points, whose coordinates and physical
        interpretation are briefly described.
	\begin{enumerate}[I.]
	  \item {\it Trivial fixed point} %-- {\bf unstable}:
         	\begin{align}
	           g_1^* &= g_2^* =0, %\quad  \forall a^*, b^*,
	           \quad
           	 \Omega  = \left[0,\, 0,\, -\varepsilon,\, -\delta \right]. 
          	\end{align}
          	This fixed point represents the free (Gaussian) FP for which all interactions are
		irrelevant and ordinary perturbation theory is applicable.
		As expected, this regime is IR stable in a region of high space dimensions
		$d>4\, (\eps<0)$ or irrelevant velocity fluctuations $\delta<0$ \cite{turbo}.
		Note that neither the coordinates nor the eigenvalues do depend on 
		fixed points' values of parameters $a$ and $b$. Two vanishing 
		eigenvalues correspond to
		a marginal plane and instead of a fixed point we actually have a whole plane.
		This a relatively common pattern seen for trivial FPs \cite{kr4,Ant00}.
%------------------------------------------------------------------------------------------------%		
 	  \item {\it Non-equilibrium thermal fluctuations}  %-- {\bf unstable}:
          	\begin{align}
               	  g_1^* & = \varepsilon \sqrt{b + 1}/3, \quad a^* = 1/4, \quad g_2^*=0, \\ % \quad  \forall \, b^*, \\ 
                  \Omega & = \left[\varepsilon,\, - \delta,\, 0 ,\, \varepsilon/3 \right].
	          \end{align}
		In this case, the velocity fluctuations are irrelevant, because
		the corresponding charge $g_2$ attains zero value. On the other
		hand coordinate $g_1$ is non-zero and
		thus this regime corresponds to a critical regime of ordinary model A 
		\cite{nav1,Zinn}.
		We observe that this regime is stable below space dimension $4$ and irrelevant
		velocity fluctuations.
%------------------------------------------------------------------------------------------------%		
	  \item  {\it Turbulent mixing of a passive scalar}  %-- {\bf unstable}:
		\begin{align}
		    \label{eq:DP_regime3a}
		    g_2^* & = \frac{24\, \delta}{\left\{  \alpha (6 + \rho_3)+5 \rho_1+\rho_2+18 \right\}}, 
		    &g_1^*& = 0, %\\ %[10pt]
		    \\
		    b^* &  =  \frac{2\,\{  \alpha \rho_3 - \rho_1 + 7 \rho_2\}}{\left\{  \alpha (6 + \rho_3)+5
		    \rho_1+\rho_2+18 \right\}},\\ % \quad \forall  a^*,\\		    		
		    \Omega & = \left[\Omega_1,\, \delta,\, \delta,\, 0 \right], 		    
		    \label{eq:DP_regime3b}
		\end{align}
		where $ \Omega_1 = {\partial \beta_{g_1}}/{ \partial g_1} |_{g = g^*}$.
		In this regime, the interaction term $\varphi^3\varphi'$ is infra-red 
		irrelevant, and the order 
		parameter $\varphi$ behaves like a passive mixture described by a diffusion 
		equation with an
		advecting term \cite{n5,n6,n7}. Turbulent mixing in this case is so strong that it
		completely destroys self-interaction of the field $\varphi$. Note that the 
		fixed points'
		coordinate of the parameter $a$ is not fixed and does not influence 
		the stability region.
%------------------------------------------------------------------------------------------------%		
	  \item  {\it Interaction of thermal and turbulent fluctuations}:
		\begin{align}
		    g_2^* & = \frac{24\, \delta}{\left\{  \alpha (6 + \rho_3)+5 \rho_1+\rho_2+18 \right\}}, 
		    \quad g_1^*,\\ 
		    b^* & =  \frac{2\,\{  \alpha \rho_3 - \rho_1 + 7 \rho_2\}}{\left\{  \alpha (6 + \rho_3)+5
		    \rho_1+\rho_2+18 \right\}}, \quad a^* = 1/4,
		    \label{eq:A_FPcoordB}
		    \\				
		   \Omega & = \left[3 \,\Omega_4,\, \delta,\, \delta,\, \Omega_4 \right], \quad \Omega_4 =
		   \left.\frac{\partial \beta_a}{ \partial a}\right|_{g = g^*  }.
		\end{align}
		An analytic expression for $g_1^*$ can be simply derived from (\ref{eq:betaA3}).
%%%%%%%%%%%%%%%%%%%%%%%%%%%%%%%%%%%%%%%%%%%%%%%%%%%%%%%%%%%%%%%%%%%%%%%%%%%%%%%%%%
%
%               CHANGED IN REVISION
%
%%%%%%%%%%%%%%%%%%%%%%%%%%%%%%%%%%%%%%%%%%%%%%%%%%%%%%%%%%%%%%%%%%%%%%%%%%%%%%%%%%		
{
	      Its explicit form can be found in Appendix~\ref{app:explicit}, see Eq.~(\ref{eq:modelAg1}).
	      }
%%%%%%%%%%%%%%%%%%%%%%%%%%%%%%%%%%%%%%%%%%%%%%%%%%%%%%%%%%%%%%%%%%%%%%%%%%%%%%%%%%
%
%               END OF REVISION
%
%%%%%%%%%%%%%%%%%%%%%%%%%%%%%%%%%%%%%%%%%%%%%%%%%%%%%%%%%%%%%%%%%%%%%%%%%%%%%%%%%%	
	\end{enumerate}	
	In order to gain an insight into properties of regime IV we have numerically found corresponding
	zeros of $\beta$-functions and then obtain eigenvalues using a diagonalization procedure.
	All of these steps are easy to incorporate in a symbolic computational program \cite{maple}.	
	 In particular,	we have
	analyzed possible regimes at asymptotic values of parameters $\rho_1, \rho_2, \rho_3$ and $\alpha $.
	We were able to obtain analytic expressions
	and they can be summarized as follows:
	\begin{enumerate}
		\item ${\alpha=0}$ -- { unstable} for any $ \rho_1, \rho_2, \rho_3$.
		\item $0<\alpha<\infty$
		\begin{enumerate}
			\item ${\rho_1 \sim \rho_2 \sim \rho_3 \rightarrow \infty}$,\\
			      {stable} for $ \alpha > 30 / (67 - 36\, \sqrt{3})$\\$\approx 6.45 $,
			\item ${\rho_1 \sim \rho_2 \rightarrow \infty, \rho_3 = 0}$, 
			{unstable} for any $ \alpha$,	
			\item ${\rho_1 \sim \rho_3 \rightarrow \infty, \rho_2 = 0}$,\\
			{stable} for $ \alpha > 5\, (2\,\sqrt{6}+5)/(25 - 2\,\sqrt{6}) \approx 2.46$,
			\item ${\rho_2 \sim \rho_3 \rightarrow \infty, \rho_1 = 0}$,\\ 
			{stable} for $\alpha > 5\, (8 \sqrt{15}+83)/(311 - 40\, \sqrt{15}) \approx 3.65$,
			\item ${\rho_1 = \rho_2 =0, \rho_3 \rightarrow \infty}$,
			{stable} for any $ \alpha$,
			\item  ${\rho_1 = \rho_3 =0, \rho_2 \rightarrow \infty}$, 
			{unstable} for any $ \alpha$	,
			\item  ${\rho_2 = \rho_3 =0, \rho_1 \rightarrow \infty}$,
			{unstable} for any $ \alpha$,
			\item ${ \rho_1 =\rho_2 = \rho_3 =0}$,\\
			{stable} for $ \alpha > 15/7 \approx 2.14$ (see \cite{an7}).
		\end{enumerate}
		\item $\alpha=\infty$ -- {stable} for any $ \rho_1, \rho_2, \rho_3$.
	\end{enumerate}
\if 0
\begin{figure}
 \centering
\resizebox{0.35\textwidth}{!}{ \includegraphics{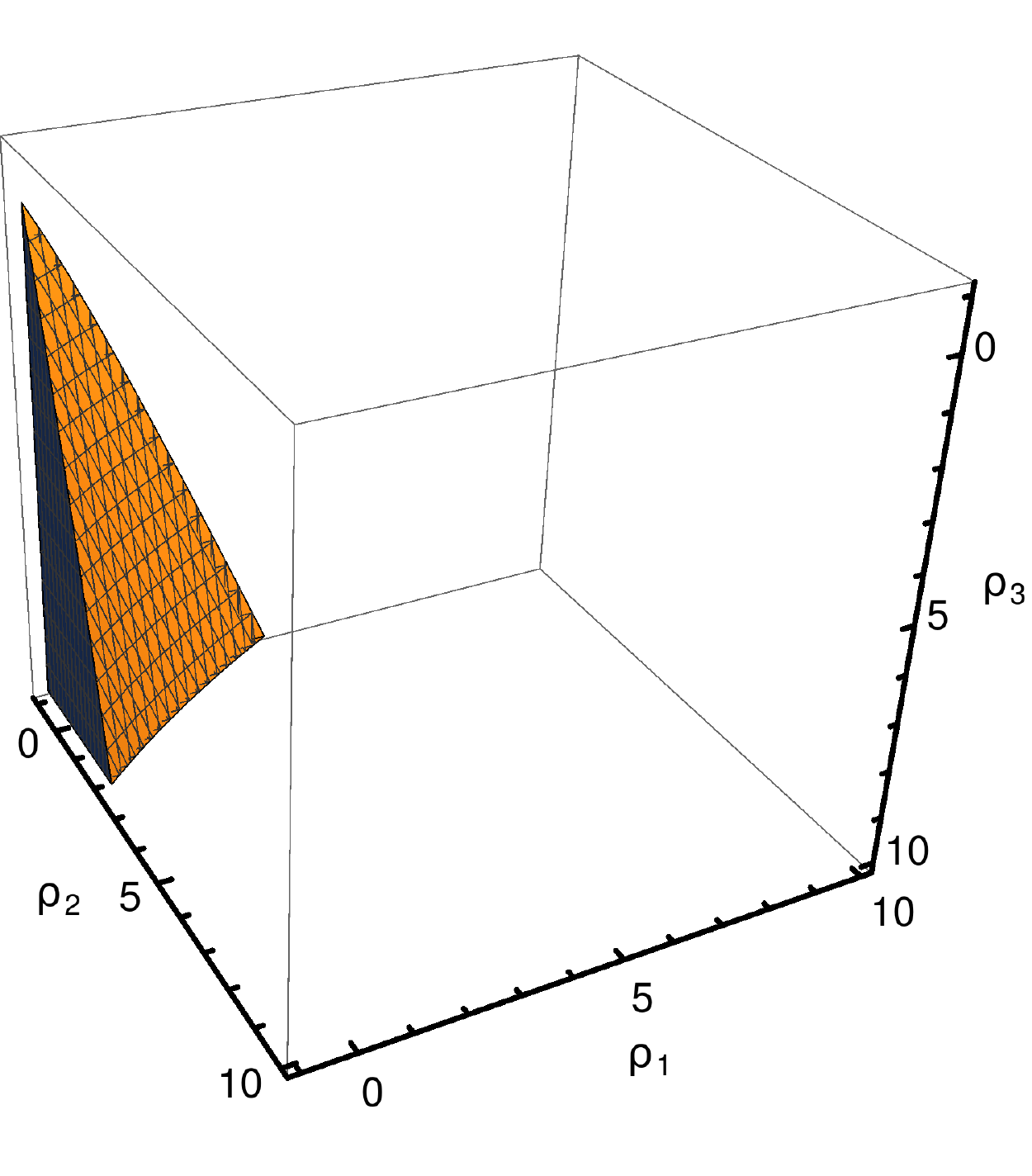} }
%\vspace{5cm}       % Give the correct figure height in cm
\resizebox{0.35\textwidth}{!}{ \includegraphics{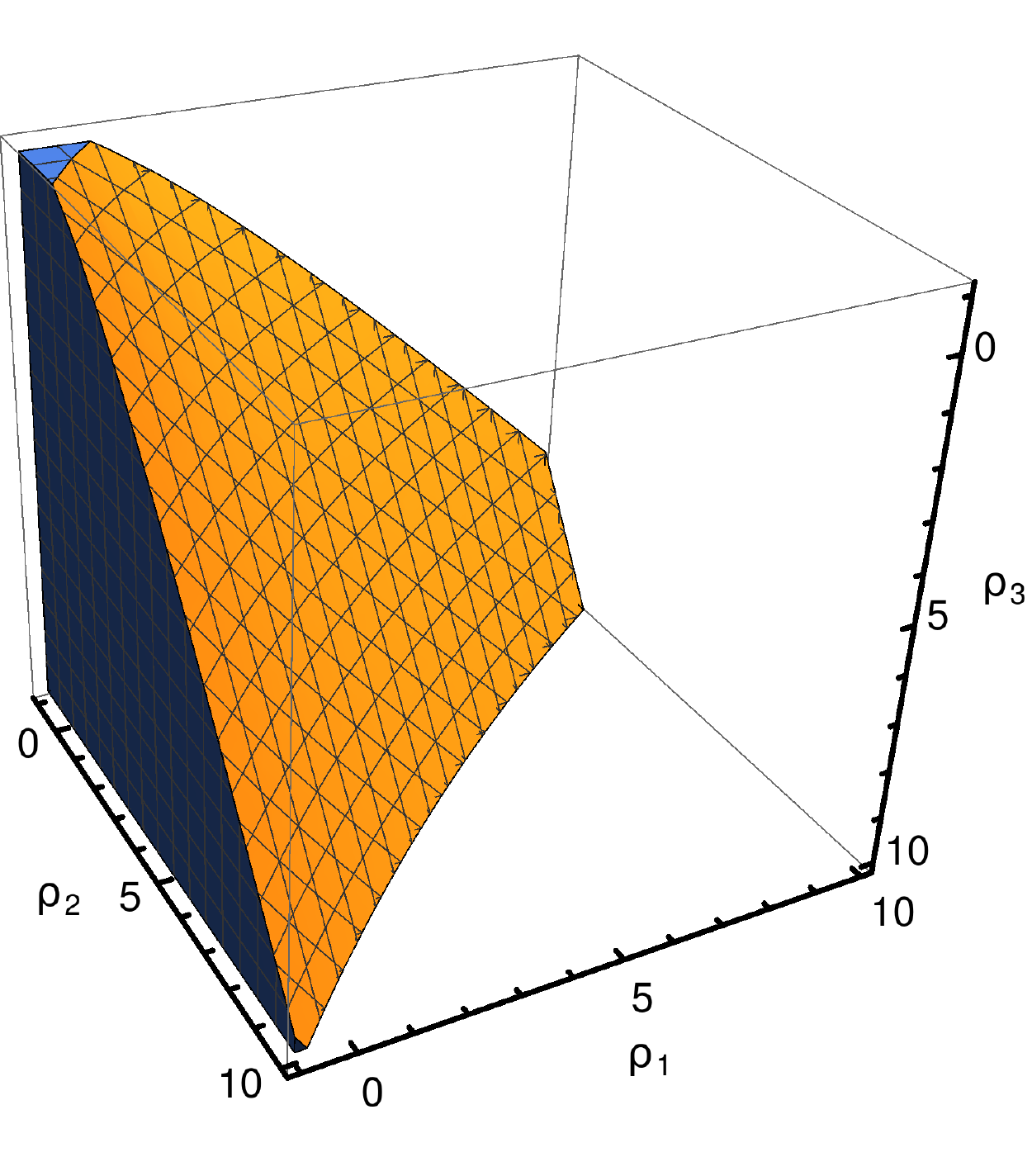} }
%\vspace{5cm}       % Give the correct figure height in cm
  \resizebox{0.35\textwidth}{!}{ \includegraphics{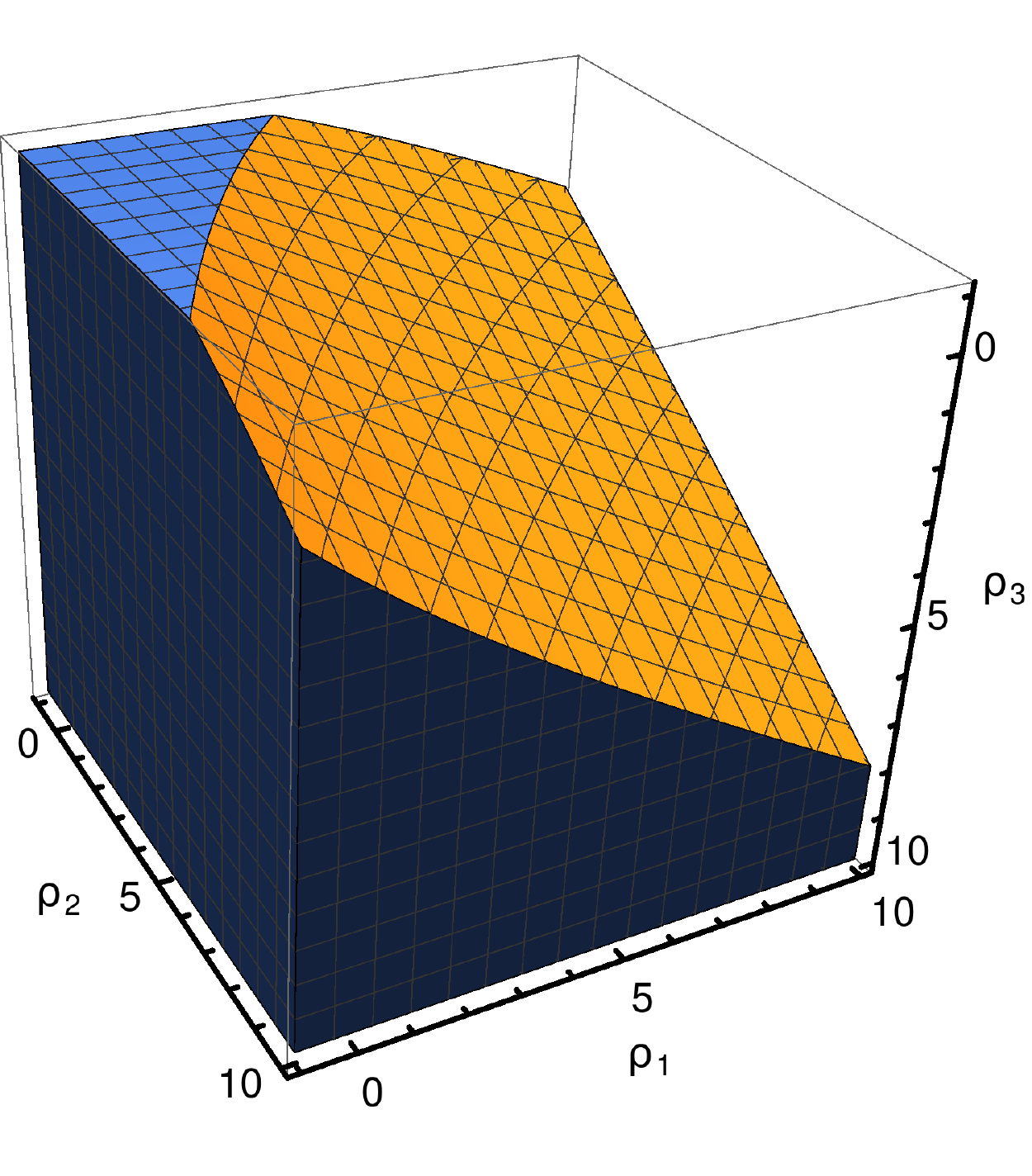} }
%\vspace{5cm}       % Give the correct figure height in cm
\caption{Stability regions for fixed point IV in model A. Dark blue
      regions stand for volumes in the three-dimensional parameter space spanned
      by $\rho_1,\rho_2$ and $\rho_3$, where the stability matrix
      has positive eigenvalues. Four different realization are depicted for 
      $\alpha=1,2.5$ and $5$. }
  \label{fig:IVa} 
\end{figure}

\fi 

\begin{figure}%
    \centering
    \subfloat[$\alpha=1$]{{\includegraphics[width=0.2\textwidth]{IVVOL1.pdf} }}%
    \qquad
    \subfloat[$\alpha=2.5$]{{\includegraphics[width=0.2\textwidth]{IVVOL25.pdf} }}%
    \qquad
    \subfloat[$\alpha=5$]{{\includegraphics[width=0.2\textwidth]{IVVOL5.pdf} }}%
    \qquad
    \subfloat[$\alpha=10$]{{\includegraphics[width=0.2\textwidth]{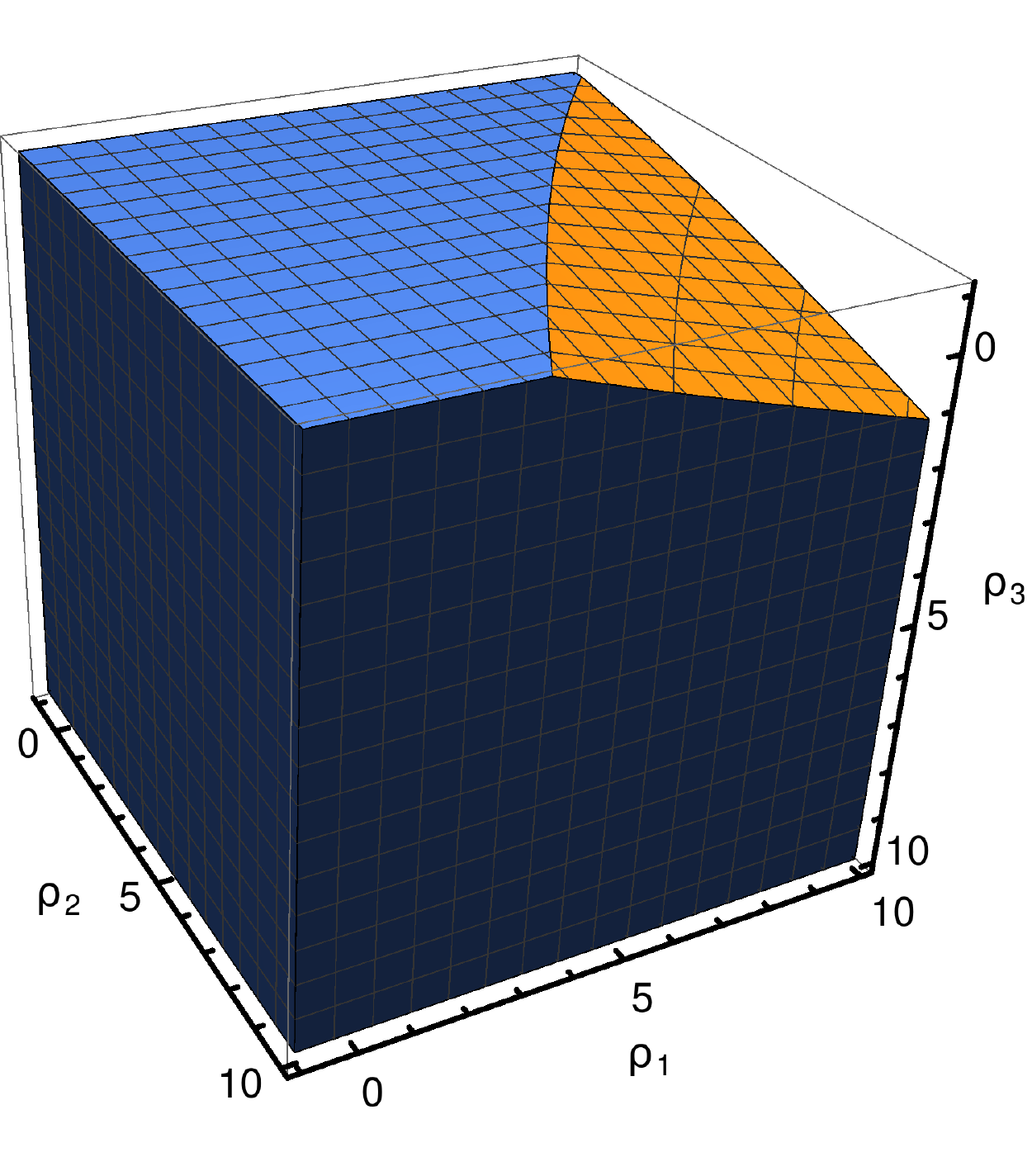} }}%
    \qquad
    \caption{Stability regions of fixed point IV in model A for different values of ``compressibility'' $\alpha$. Dark blue
      regions stand for volumes in the three-dimensional parameter space spanned
      by $\rho_1,\rho_2$ and $\rho_3$, where the stability matrix
      has positive eigenvalues.}%
      \label{fig:IVa}
    \end{figure}

	A cumbersome explicit expression for eigenvalue $\Omega_4$ is not provided here. 
	Instead, we consider a smooth section of the hypersurface determined by the equation 
	\begin{equation*}
	  \Omega_4(\rho_1, \rho_2, \rho_3, \alpha)=0
	\end{equation*}
	at given value of compressible parameter $\alpha$. Points belonging 
	to the volume 
	\begin{equation}
	  {\mathcal B}= \{ \rho_i >-1\, (i=1,2,3), \alpha>0:
	  \Omega_4(\rho_1, \rho_2, \rho_3, \alpha)>0\}
	\end{equation}
	determine the IR stability region.
	 Volume  $ |{\mathcal B}|$ for four typical values of $\alpha$ is depicted in
	Fig. \ref{fig:IVa}.  $ |{\mathcal B}|$  vanished fo the incompressible fluid
	and increases indefinitely 
         with compressible parameter $\alpha$. A change of
	the anisotropy parameters ($\rho_1, \rho_2, \rho_3$)
	at fixed $\alpha$ may affect the IR stability
	of the fixed point according to Fig. \ref{fig:IVa}.
	Also, it is worth mentioning that anisotropy of a
	mixing flow does not affect the type of 
	critical asymptotics in both incompressible ($\alpha=0$) and infinitely compressible 
	($\alpha=\infty$) case. For the finite values of compressibility parameter
	$\alpha$ the anisotropy is an
	important factor in establishing scaling regimes. This clearly displays the failure of the
	Kolmogorov hypothesis (that the anisotropy dies out in the inertial range) in the case 
	of critical fluids. In a vicinity of the point $\rho_1 = \rho_2 = \rho_3 =0$ at
	fixed $\alpha$ weak anisotropy does not change IR stability.

	Let us summarize main findings about obtained scaling regimes in model A that
	correspond to fixed point IV:
\begin{enumerate}[(i)]
  \item Regime $1$. Incompressibility of the system entails
	that fluid is far from a critical or phase transition
	point. Hence, long-wave fluctuations of the order parameter do not determine large-scale
	behavior. They are suppressed by the transversal turbulent fluctuations, which play a 
	crucial role in the realization of IR scaling.
  \item Regime $2$ without the inclusion of case $(e)$. In these cases
	the order parameter fluctuations are
	still weaker in comparison to the velocity ones.
	We can observe (items $(b),(f),(g)$) that the
	transversal part ($\rho_1, \rho_2$) of the turbulent fluctuations inhibits the role
	of iso\-tro\-pic ($\rho_3=0$) longitudinal components for all permissible
	$ \alpha$.  On the other hand, in regimes
	$(a)$,$(c)$,$(d)$, and $(h)$ the transversal and longitudinal parts become
	equally relevant. 
  \item Regime $3$ and the case $(e)$. These regimes correspond to a strong 
        compressible fluid. It has been
	mentioned above that compressibility grows very fast in the vicinity of the 
	critical point,
	where fluctuations of the order parameter could not be neglected.
\end{enumerate}

%------------------------------------------------------------------------------------------------------------    
{\subsection{Model DP} \label{sec:model_DP}}
%------------------------------------------------------------------------------------------------------------
Let us now analyze scaling regimes of the DP model~(\ref{eq:SDP}). This is
given by fixed points' solution
of the set of equations (\ref{eq:betaG1})-(\ref{eq:betaG4}).

\begin{enumerate}[I.]
 \item {\it Trivial (Gaussian) fixed point} %-- {\bf unstable}:
	\begin{align}
	  g_1^* &= g_2^* =0, \quad % \forall a^*, b^*, \\
	  \Omega = \left[0,\, 0,\, -\varepsilon,\, -\delta \right]. 
	\end{align}	
	Similarly to the case of model A, this fixed point represents the free FP for which all interactions are
		irrelevant and values of parameters $a$ and $b$ do not change its 
		stability region. As expected this fixed point is stable above
		space dimension $d_c=4$  and negligible velocity fluctuations.
  %------------------------------------------------------------------------------------------------%		
 \item {\it Non-equilibrium thermal fluctuations} %-- {\bf unstable}:
	\begin{align}
	  g_1^* & = 2\, \varepsilon \sqrt{b+1}/3 , \quad a^* = 1/2, \quad g_2^*=0, \\ %  \forall \, b^*,\\
	  \Omega &  = \left[\varepsilon,\, \varepsilon/12 - \delta,\, 0,\, \varepsilon/6 \right].
	\end{align}
	In this regime, the stochastic velocity fluctuations are irrelevant and
	charge $b$ is not fixed.
	This regime corresponds to a pure DP model.  For more details and 
	discussions of this regime  see \cite{tauber}.
  %------------------------------------------------------------------------------------------------%	
 \item {\it Turbulent mixing of a passive scalar} %-- {\bf unstable}:
	\begin{align}
	   g_2^* & = \frac{24\, \delta}{\left\{  \alpha (6 + \rho_3)+5 \rho_1+\rho_2+18 \right\}},
	   &g_1^*& = 0, %\\
	   \\
           b^* & =  \frac{2\,\{  \alpha \rho_3 - \rho_1 + 7 \rho_2\}}{\left\{  \alpha (6 + \rho_3)+5 
           \rho_1+\rho_2+18 \right\}}, % \quad \forall a^*,\\
           \\
           %------------------------------------------------------------------------------------
	   \Omega & = \left[\Omega_1,\, \delta,\,  \delta,\, 0 \right], 	   
	\end{align}
	where $ \Omega_1 = {\partial \beta_g}/{ \partial g} |_{ \kappa = \kappa^*}$.
	This fixed point  is similar to regime III of model A given by coordinates 
	(\ref{eq:DP_regime3a}-\ref{eq:DP_regime3b}). Once again DP non-linearities are
	completely suppressed by vigorous turbulent fluctuations. 
%%%%%%%%%%%%%%%%%%%%%%%%%%%%%%%%%%%%%%%%%%%%%%%%%%%%%%%%%%%%%%%%%%%%%%%%%%%%%%%%%%
%
%               ADDED IN REVISION
%
%%%%%%%%%%%%%%%%%%%%%%%%%%%%%%%%%%%%%%%%%%%%%%%%%%%%%%%%%%%%%%%%%%%%%%%%%%%%%%%%%%
{
	Similarly, the parameter $a$ is not fixed. 
	}
%%%%%%%%%%%%%%%%%%%%%%%%%%%%%%%%%%%%%%%%%%%%%%%%%%%%%%%%%%%%%%%%%%%%%%%%%%%%%%%%%%
%
%               END OF REVISION
%
%%%%%%%%%%%%%%%%%%%%%%%%%%%%%%%%%%%%%%%%%%%%%%%%%%%%%%%%%%%%%%%%%%%%%%%%%%%%%%%%%%
	
 %------------------------------------------------------------------------------------------------%	
 \item  {\it Interaction of thermal and hydrodynamic fluctuations }	
	\begin{align}
	   %g_2^* & = \frac{24\, \delta}{\left\{  \alpha (6 + \rho_3)+5 \rho_1+\rho_2+18 \right\}}, 
	   %& g_1^*&, \\
	   g_1^*&, \quad g_2^*, \quad a^* = 1/2, \\
	   %------------------------------------------------------------------------------------
           b^* & =  \frac{2\,\{  \alpha \rho_3 - \rho_1 + 7 \rho_2\}}{\left\{  \alpha (6 + \rho_3)
           +5 \rho_1+\rho_2+18 \right\}},
           \label{eq:DP_FPcoordB}
           \\
           %------------------------------------------------------------------------------------
	    \Omega & = \left[2 \,\Omega_4,\, \Omega_2,\, \Omega_2,\, \Omega_4 \right], 	    
	\end{align}
	where $\Omega_4 = {\partial \beta_a} / { \partial a} |_{\kappa = \kappa^* }$, 
	    $\Omega_2 = {\partial \beta_{g_2}} / { \partial g_2} |_{\kappa = \kappa^*}$.
	One can show that
	$\Omega_2 >0$ for $\forall \rho_1,\, \rho_2, \, \rho_3,\, \alpha$.
	Expressions for
	 fixed point's coordinates  $g_1^*$ and $g_2^*$ can be derived from Eqs.~(\ref{eq:betaG1})-(\ref{eq:betaG4}), and
%%%%%%%%%%%%%%%%%%%%%%%%%%%%%%%%%%%%%%%%%%%%%%%%%%%%%%%%%%%%%%%%%%%%%%%%%%%%%%%%%%
%
%               CHANGED IN REVISION
%
%%%%%%%%%%%%%%%%%%%%%%%%%%%%%%%%%%%%%%%%%%%%%%%%%%%%%%%%%%%%%%%%%%%%%%%%%%%%%%%%%%	
{
	      their explicit form can be found in Appendix~\ref{app:explicit}, see Eq.~(\ref{eq:modelDPg1}).
	      }
%%%%%%%%%%%%%%%%%%%%%%%%%%%%%%%%%%%%%%%%%%%%%%%%%%%%%%%%%%%%%%%%%%%%%%%%%%%%%%%%%%
%
%               END OF REVISION
%
%%%%%%%%%%%%%%%%%%%%%%%%%%%%%%%%%%%%%%%%%%%%%%%%%%%%%%%%%%%%%%%%%%%%%%%%%%%%%%%%%%

  %------------------------------------------------------------------------------------------------%	
\end{enumerate}	

Possible regimes at asymptotic
values of ($\rho_1, \rho_2, \rho_3, \alpha $) for fixed point IV can be summarized as follows:
	\begin{enumerate}
		\item ${\alpha=0}$ -- { unstable} for %$\forall \rho_1, \rho_2, \rho_3$
		      any $ \rho_1, \rho_2, \rho_3$
		\item $0<\alpha<\infty$
		\begin{enumerate}
			\item ${\rho_1 \sim \rho_2 \sim \rho_3 \rightarrow \infty}$,\\ 
			      {stable} for $ \alpha > 30 \,(2 + \sqrt{3}) / (14 - 5\, \sqrt{3})  \approx 20.96 $
			\item ${\rho_1 \sim \rho_2 \rightarrow \infty, \rho_3 = 0}$,
			      {unstable} for any %$\forall \alpha$
			      $\alpha$	
			\item ${\rho_1 \sim \rho_3 \rightarrow \infty, \rho_2 = 0}$,\\ 
			      {stable} for $ \alpha > (505 + 60\,\sqrt{70})/121 \approx 8.32$
			\item ${\rho_2 \sim \rho_3 \rightarrow \infty, \rho_1 = 0}$,\\
			      {stable} for $\alpha > 30 \,(2 + \sqrt{3}) / (14 - 5\, \sqrt{3})  \approx 20.96 $
			\item ${\rho_1 = \rho_2 =0, \rho_3 \rightarrow \infty}$, 
			      {stable} for %$\forall \alpha$
			      any $\alpha$
			\item  ${\rho_1 = \rho_3 =0, \rho_2 \rightarrow \infty}$,
			      {unstable} for %$\forall \alpha$
			      any $\alpha$
			\item  ${\rho_2 = \rho_3 =0, \rho_1 \rightarrow \infty}$,
			      {unstable} for % $\forall \alpha$
			      any $ \alpha$
			\item ${ \rho_1 =\rho_2 = \rho_3 =0}$, 
				{stable} for $ \alpha > 5$ (see \cite{an7})
		\end{enumerate}
		\item $\alpha=\infty$ -- {stable} for % $\forall \rho_1, \rho_2, \rho_3$.
		any  $\rho_1, \rho_2, \rho_3$.
        
	\end{enumerate}

	\if 0
\begin{figure}
 \centering
\resizebox{0.35\textwidth}{!}{ \includegraphics{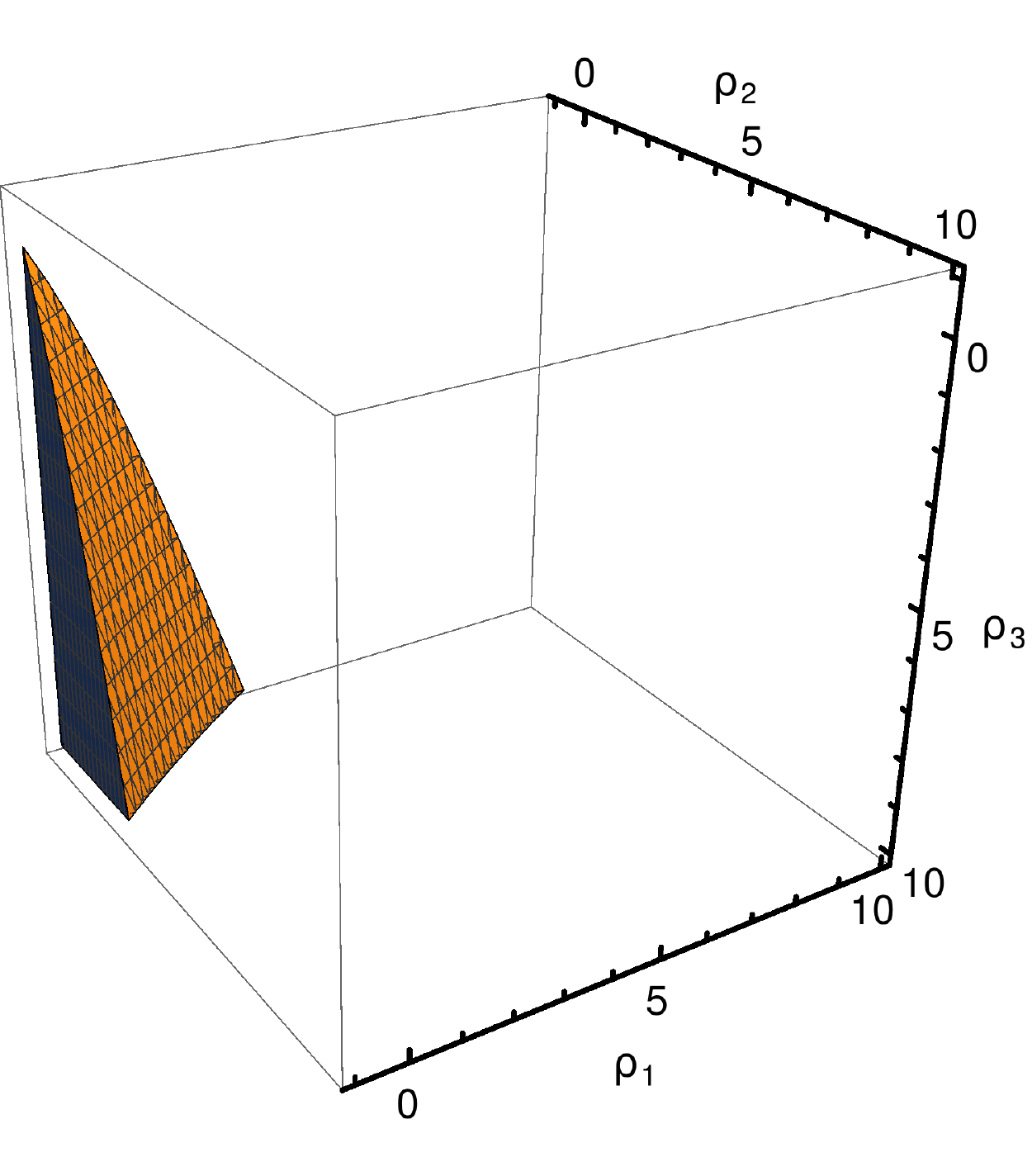} }
%\vspace{5cm}       % Give the correct figure height in cm
\resizebox{0.35\textwidth}{!}{ \includegraphics{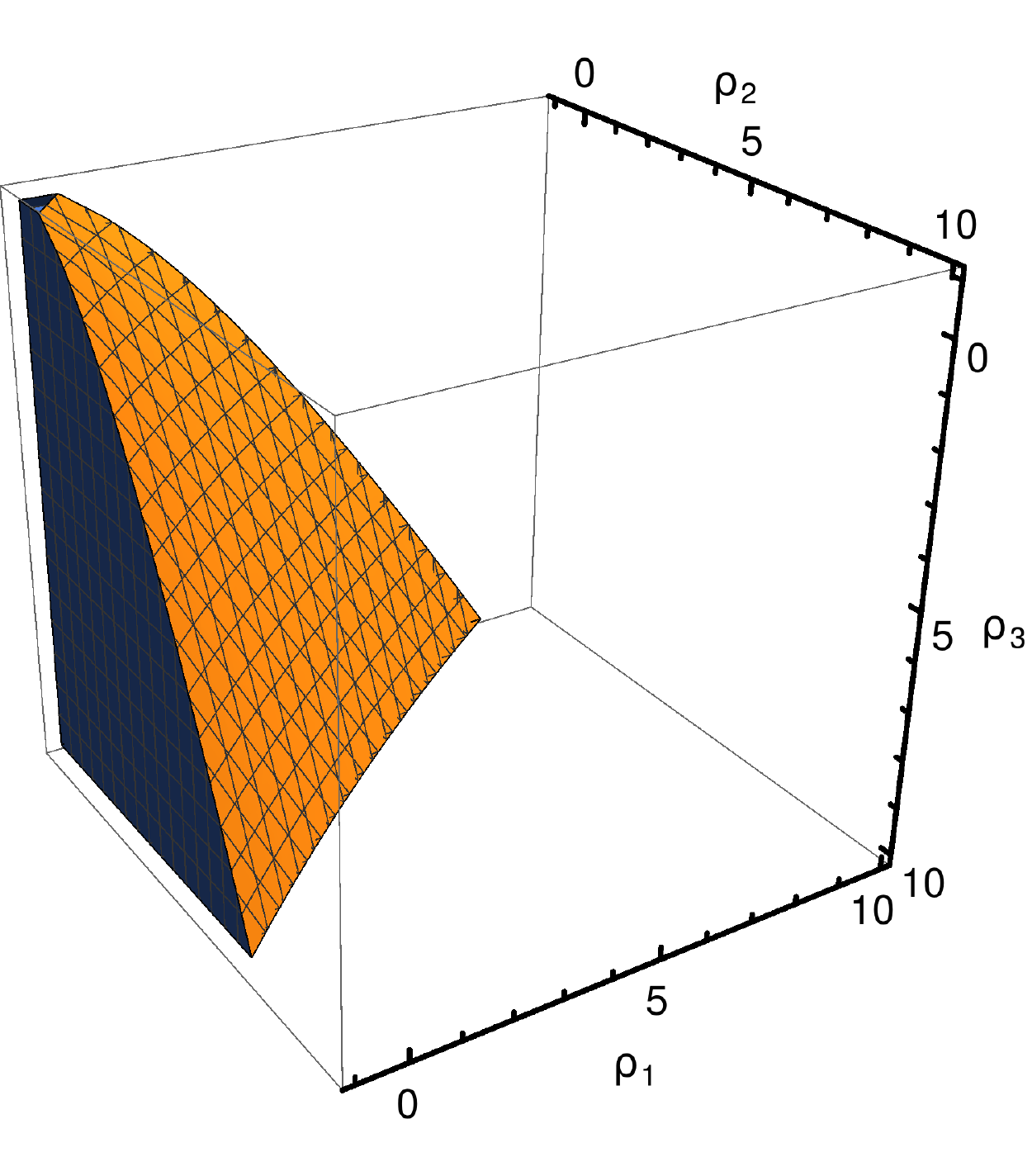} }
%\vspace{5cm}       % Give the correct figure height in cm
  \resizebox{0.35\textwidth}{!}{ \includegraphics{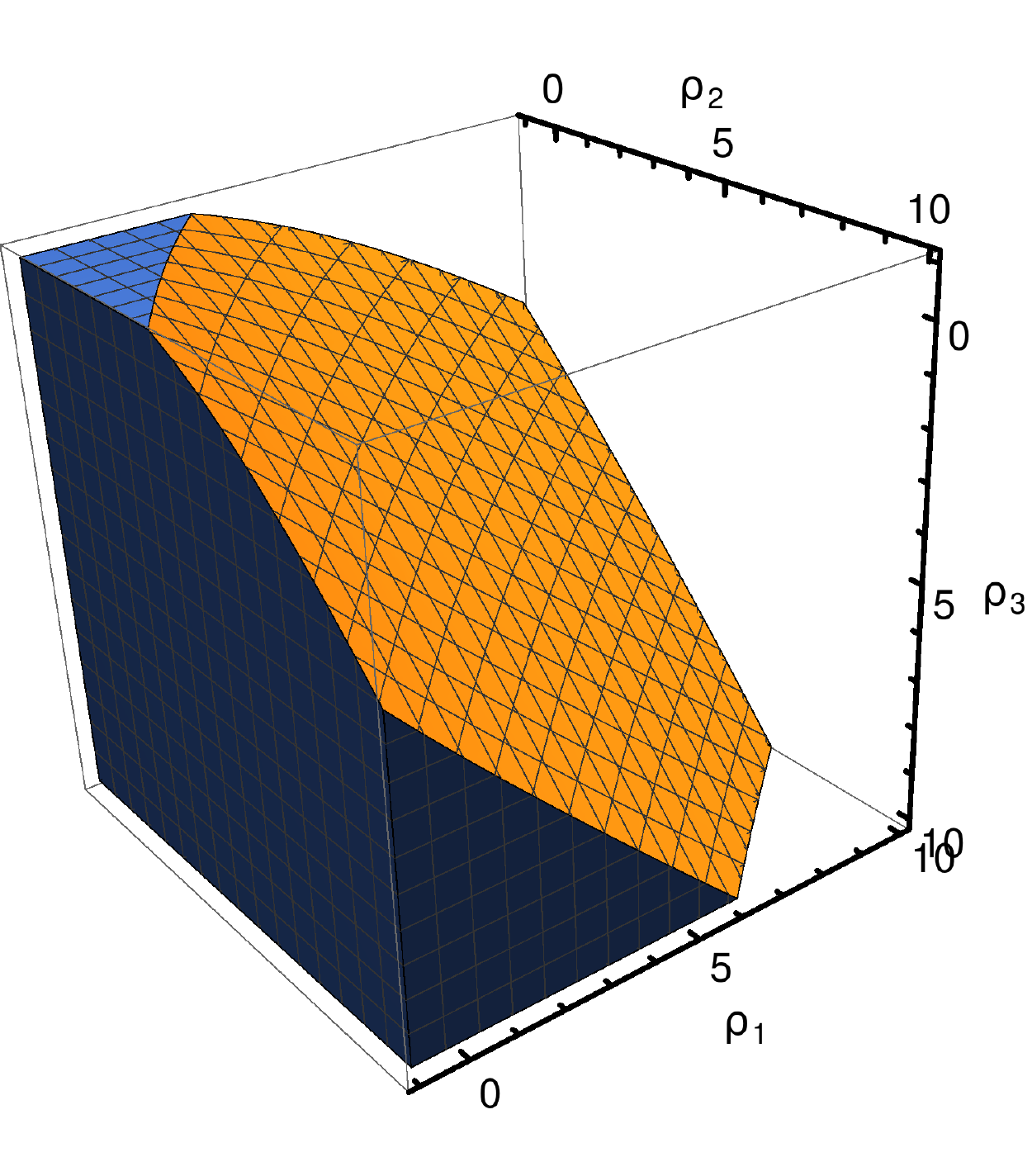} }
%\vspace{5cm}       % Give the correct figure height in cm
\caption{Stability regions for fixed point IV for  model DP. Dark blue
 		regions denotes volumes in three-dimensional parameter space spanned
 		by $\rho_1,\rho_2$ and $\rho_3$, where the stability matrix
 		has positive eigenvalues. Four different realization are depicted for 
 		$\alpha=2.5, 5 $ and $10$}
  \label{fig:IV} 
\end{figure}
\fi

\begin{figure}%
    \centering
    \subfloat[$\alpha=2.5$]{{\includegraphics[width=0.2\textwidth]{G4A25.pdf} }}%
    \qquad
    \subfloat[$\alpha=5$]{{\includegraphics[width=0.2\textwidth]{G4A5.pdf} }}%
    \qquad
    \subfloat[$\alpha=10$]{{\includegraphics[width=0.2\textwidth]{G4A10.pdf} }}%
    \qquad
    \subfloat[$\alpha=20$]{{\includegraphics[width=0.2\textwidth]{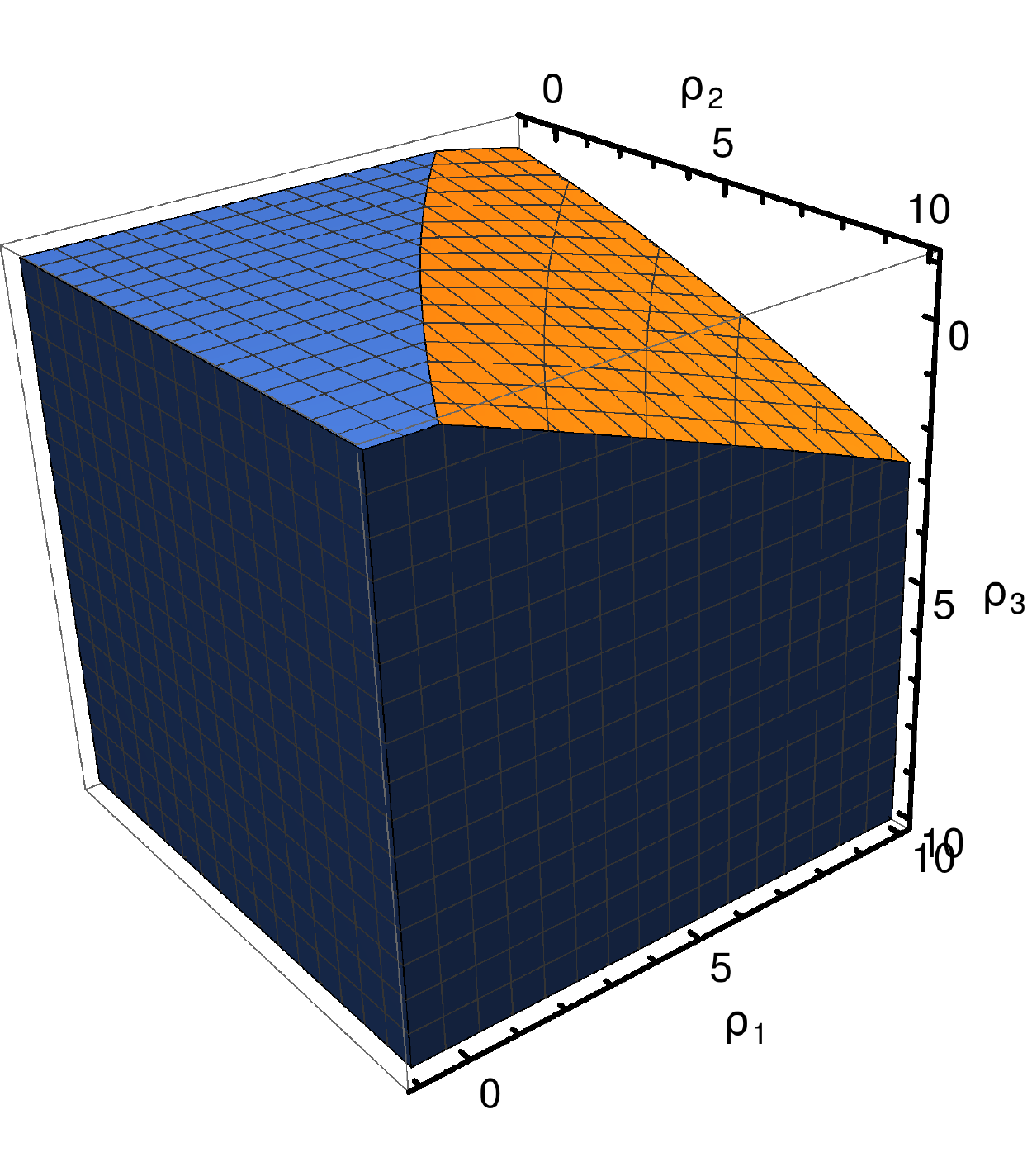} }}%
    \qquad
    \caption{Stability regions of fixed point IV in model DP for different values of ``compressibility'' $\alpha$. Dark blue
      regions stand for volumes in the three-dimensional parameter space spanned
      by $\rho_1,\rho_2$ and $\rho_3$, where the stability matrix
      has positive eigenvalues.}%
       \label{fig:IV} 
    \end{figure}

The volume 
\begin{equation}
  {\mathcal B}= \{ \rho_i >-1 (i=1,2,3), \alpha>0:\Omega_4(\rho_1, \rho_2, \rho_3, \alpha)>0\}  
\end{equation}
 of IR stable points is displayed
	 in Fig.~{\ref{fig:IV}}. An analysis of this regime reveals that turbulence has similar 
	 qualitative effects on the behavior of both model A and model DP despite a different
	 nature and origin of systems they describe.

Let us consider the stability of the regimes in a case when anisotropy function  $c(\varphi)>0$
(\ref{eq:rot}) is an arbitrary one.   Direct calculations lead to the following relation
at $\eps=1$, $\delta_K=4/3$:
\begin{equation}
\label{eq:o4}
  \Omega_4 \sim \int_{0}^{1} \left[ \frac{3}{\sqrt{b^* t + 1}} - \frac{5}{b^* + 4}\right] c(t)\, 
  (1-t)^{1/2} \, t^{-1/2} {\mathrm{d}} t,
\end{equation}
where $t = \cos(\varphi)^2$. There is no doubt that in the physical region
 $b^* >0$ eigenvalue $\Omega_4$
is  positive. We can conclude that  regardless of anisotropy intensity and structure of function
$c(\varphi)$ turbulent fluctuations do not change a type of IR regime in the infinitely
compressible system. They  affect only the values of critical exponents.
%------------------------------------------------------------------------------------------------------------
{\section{ Critical exponents} \label{sec:phys}}
%------------------------------------------------------------------------------------------------------------
 Green functions of the theory contain an essential physical information about
 a state of the system. In particular, in the percolation process the radius of gyration $R(t)$ for
the spreading cloud of the agent that started from the origin at time $t'=0$
can be expressed through the linear response function
\begin{align}
  {R^2(t)} & = \frac{\int |\mx|^2	\, G(|\mx|, t) \, 
  {\mathrm{d}}^d x }{2 d \int \, G(| \mx |, t) \, {\mathrm{d}}^d x  }, 
  \label{eq:radius}
  \\
   G(| \mx |, t) & = \langle \varphi(t, \mx ) \varphi'(0, {\bm 0})  \rangle.
\end{align}
The RG analysis yields the following IR asymptotic solution for the response function
\begin{equation}
  G(|\mx|, t) = |\mx|^{- \Delta_{\phi} - \Delta_{\phi'}} \, f(| \mx |\, t^{-1/\Delta_{\omega}}),
\end{equation}
where $f$ is a scaling function  (it also depends on $\alpha, \rho_i$ and the angle between
$\mx$ and $\mn$), $\Delta_{\varphi} = d/2 + \gamma^*_{\varphi}, \Delta_{\varphi'} = d/2 + \gamma^*_{\varphi'} $
are the scaling dimensions of fields, and $\Delta_{\omega} = 2 - \gamma^*_{\lambda}$  is the scaling 
dimension of frequency or a dynamical critical exponent \cite{nav1}. The additive $\gamma_j$ is 
 a so-called anomalous dimension corresponding to a parameter or field $j$. 
 Such $\gamma_j$  originates during the renormalization  process. Symbol $^*$ denotes a value 
 of anomalous dimensions at a fixed point. As the 
 result, we get a power-law asymptotic expression for $R^2(t)$
\begin{equation}
  R^2(t) \sim  t^{z_s}, \quad z_s = \frac{2}{\Delta_{\omega}}.
\end{equation}
 %%%%%%%%%%%%%%%%%%%%%%%%%%%%%%%%%%%%%%%%%%%%%%%%%%%%%%%%%%%%%%%%%%%%%%%%%%%%%%%%%%
%
%               ADDED IN REVISION
%
%%%%%%%%%%%%%%%%%%%%%%%%%%%%%%%%%%%%%%%%%%%%%%%%%%%%%%%%%%%%%%%%%%%%%%%%%%%%%%%%%%
{
Let us note that in contrast to \cite{turbo}, radius of gyration (\ref{eq:radius}) is defined in a different way.
The denominator in this expression corresponds  to a number of active particles generated by a single seed at initial time. Thus,
 Eq. (\ref{eq:radius}) represents the mean squared radius on one active particle at a given time.
 }
%%%%%%%%%%%%%%%%%%%%%%%%%%%%%%%%%%%%%%%%%%%%%%%%%%%%%%%%%%%%%%%%%%%%%%%%%%%%%%%%%%
%
%               END OF REVISION
%
%%%%%%%%%%%%%%%%%%%%%%%%%%%%%%%%%%%%%%%%%%%%%%%%%%%%%%%%%%%%%%%%%%%%%%%%%%%%%%%%%%

The scaling behavior for a number of active particles
\begin{equation}
  {N(t)} = \int \, G(| \mx|, t) \, {\mathrm{d}}^d x
\end{equation}
is given by the asymptotic relation
\begin{equation}
  N(t) \sim  t^{\theta_s}, \quad \theta_s = -\frac{(\gamma^*_{\varphi} + \gamma^*_{\varphi'})}{\Delta_{\omega}}.
\end{equation}
A one-loop RG calculation leads to the expressions
\begin{align}
 \label{eq:gamma}
  \gamma_{\varphi} + \gamma_{\varphi'} & = - \frac{g_1}{4\, \sqrt{\beta+1}},
  \\
  \label{eq:gamma1}
  \gamma_{\lambda} & =  \frac{g_2}{24} \left\{  \alpha (6 + \rho_3)+5 \rho_1+\rho_2+18 \right\} 
  + \frac{g_1}{8 \sqrt{b+1}}.
\end{align}
Using  these functions and coordinates of the fixed point IV one can obtain the exponents 
$z_s$ and $\theta_s$ at given values of the parameters $(\alpha, \rho_1, \rho_2, \rho_3)$ in
the form of $(\epsilon,\delta)$-expansion up to the first order. Let us consider a case of 
weak anisotropy $\rho_1 \sim \rho_2 \sim \rho_3 \ll 1$ at $\varepsilon=1, \delta=4/3$ for  
regime IV. Then the exponent $\theta_s$ can be represented as a segment of power series in $\rho_j$
\begin{align}
  \label{eq:SS}
  \theta_s = \frac{\alpha -5}{6\,(2 \alpha+ 5)} - \frac{5 \alpha\, (\rho_1 + \rho_2 -
  \rho_3)}{8 \,(2 \alpha + 5)^2} + {\mathcal{O}}(\rho_j^2),
\end{align}
and $z_s = 2/(2 - \delta) = 3$. This relation is universal, i.e. it does depend neither on large-scale
anisotropy nor compressibility of the medium. In the previous section, it has been established that 
regime IV is unstable at small values of $\alpha$ and stable
 at its large values.   Therefore, it is reasonable to consider expression (\ref{eq:SS}) at large $\alpha$
 (formally for $\alpha \gg 5/2$)
\begin{align}
\label{eq:SSS}
  \theta_s = \frac{1}{12} - \frac{5 \, (4+ \rho_1 + \rho_2 - \rho_3)}{32\, \alpha} + {\mathcal{O}}(\alpha^{-2}).
\end{align}
The term  $(4+ \rho_1 + \rho_2 - \rho_3)$ is positive, at least in the range $|\rho_j| \leqslant 1$.
 Thus corrections to the exponent $\theta_s$ due to weak anisotropy and  finite compressibility  
reduce values of $ \theta_s$. At this point we can conclude that  number  $N(t)$  of active 
 particles grows slower than in the case of isotropic and incompressible systems.  

The IR stable regime III leads to the exact qualitative outcomes $z_s = 2/(2- \delta)= 3,
\, \theta_s = 0$, this is consistent with known Richardson's 4/3 power 
law of turbulent diffusion  $ \partial_t R^2(t) \sim R^{4/3}(t) $  \cite{turbo}.

%------------------------------------------------------------------------------------------------------------
\section{\label{sec:concl}Conclusion}
%------------------------------------------------------------------------------------------------------------

 An experimental study of critical fluids in terrestrial-like environment faces influence of Earth
 gravity inducing a distinguished direction  due to compressibility under the hydrostatic
 pressure. Systems are subject to mixing, shaking and other hydrodynamic effects in
 laboratory conditions. 
 Our analysis has
  been focused on the investigation of possible stable regimes in anisotropic turbulently
  moving critical fluids described by the A and DP effective models. The renormalization 
  group analysis within the framework of the one-loop approximation allows us to draw the
  following conclusions and describe the major findings
\begin{enumerate}[(i)]
  \item At a qualitative level both model A and model DP coupled to the Kraichnan model
	manifest the same scaling behavior. 
  %----------------------------------------------------------------------------------------------%	
  \item Close to criticality the effect of anisotropy could be relevant for establishing 
	scaling behavior. This result could call into a question well-established and universal 
	assumptions that large-scale anisotropy is irrelevant in the inertial range. A critical 
	liquid remains sensitive to a distinguished direction ${\mn}$, what implies
	that the Kolmogorov
	hypothesis does not apply to a critical fluid.
  %----------------------------------------------------------------------------------------------%		
  \item Within the considered models,  regime IV in a strong compressible system is less
	sensitive to the anisotropy of a fully developed turbulent flow. The existence of large 
	scale anisotropy does not affect its stability at the limit values of $\alpha = \infty$ 
	(see Sec.~IV, item (iii)). In the infinitely (formally) compressible system, in which
	longitudinal
	velocity fluctuations prevail due to large values of bulk viscosity, regimes stability
	also does not depend on the specific form of the function $c(\varphi)$, see Eq.~(\ref{eq:o4}).
	At the same time, quantitative values of critical exponents are not universal. 
	In fact, they are determined by function $c(\varphi)$.
%----------------------------------------------------------------------------------------------%	
  \item In the intermediate range of $\alpha$, weak (linear in $\rho_j$) anisotropy  fluctuations
	do not change the regimes. An increase of anisotropy is accompanied by the complex pattern
	of 
	interaction between transversal and longitudinal turbulent fluctuations and the
	long-wa\-ve\-length
	critical
	fluctuations of the order parameter, as depicted in Figs.~\ref{fig:IVa}, and~\ref{fig:IV}. 
	Here, anisotropy plays 
	a crucial role imposing lower limits  on the value of  $\alpha$ (on the ``compressibility 
	degree'',
	see Secs.~\ref{sec:model_A}, and \ref{sec:model_DP} and item
	denoted as IV therein). This result  is in agreement what we
	can physically anticipate: 
	order parameter
	fluctuations are strong in the vicinity of the critical point, where compressibility is
	large enough. 
  %----------------------------------------------------------------------------------------------%		
  \item Having calculated the anomalous dimensions (\ref{eq:gamma}), and (\ref{eq:gamma1}), we have
  estimated
	the critical exponents at given $\rho_j$ and $\alpha$ (\ref{eq:SS}), and (\ref{eq:SSS}).  The 
	compressibility 
	and anisotropy alter values of the critical exponent.  Whereas compressibility reduces
	the magnitude
	of the critical exponents, according to the expression (\ref{eq:SSS}), the effect of
	anisotropy is more 
	subtle and twofold, depending on the sign of the value $\rho_1+\rho_2 - \rho_3$.     
\end{enumerate}
      
 The effect of compressibility on critical behavior of an isotropic system  was revealed by the
 authors of work~\cite{an7}, where a new scaling regime IV has been identified. However, our 
 analysis shows that not only compressibility but also large-scale anisotropy may control
 the stability of the critical regimes and alter the quantitative results.  
 
%-----------------------------------------------------------------------------------------------------------
\section*{Acknowledgments}
%-----------------------------------------------------------------------------------------------------------
The authors are thankful M. Yu. Nalimov for many fruitful and inspiring discussions.
G. Kalagov is grateful for the support provided by the VVGS grant 2018-803  of PF UPJS.  
The work was supported by VEGA grant No. 1/0345/17 of the Ministry of Education, Science, 
Research and Sport of the Slovak Republic,
 the grant of the Slovak Research and Development Agency under the contract No. APVV-16-0186.
%-----------------------------------------------------------------------------------------------------------
%%%%%%%%%%%%%%%%%%%%%%%%%%%%%%%%%%%%%%%%%%%%%%%%%%%%%%%%%%%%%%%%%%%%%%%%%%%%%%%%%%
%
%               ADDED IN REVISION
%
%%%%%%%%%%%%%%%%%%%%%%%%%%%%%%%%%%%%%%%%%%%%%%%%%%%%%%%%%%%%%%%%%%%%%%%%%%%%%%%%%%
\appendix
\section{Coordinates of fixed points}
\label{app:explicit}
{
Here, we present explicit expressions for coordinates of nontrivial fixed points. First,
 at nontrivial fixed point IV of model A charge $g_1$ has the following coordinate 
 \begin{equation}
  g_1^*=\frac {\sqrt {b+1}}{3 A} \left(A \, \varepsilon -2 A \, \delta+18 B\,\delta \right).
  \label{eq:modelAg1}
\end{equation}
At nontrivial fixed point IV of model DP fixed points' values of charges $g_1$ and $g_2$ are given by the following expressions
\begin{align}
  g_1^* & = \frac {4 \sqrt {b+1}}{A + 9 B} \left(A \, \varepsilon -2 A \, \delta+18 B\,\delta \right), 
  \label{eq:modelDPg1} \\
  g_2^* & = \frac{12 \,(4\,\delta-\varepsilon)}{A+9 B}.
  \label{eq:modelDPg2}
\end{align}
In Eqs. (\ref{eq:modelAg1})-(\ref{eq:modelDPg2})  we have used the following abbreviations
\begin{align}
  A & =  \alpha  (6 + \rho_3) + 5 \rho_1  +  \rho_2 +  18, \nonumber \\
  B & =  \frac{\alpha}{ {b^*}^2} \left[ 2 (\sqrt{b^*+1}-1) (b^* - \rho_3) + b^* \rho_3 \right]. \nonumber
\end{align} 
The fixed point value $b^*$ is given by Eq.~(\ref{eq:A_FPcoordB}) or Eq.~(\ref{eq:DP_FPcoordB}), respectively.
}
%%%%%%%%%%%%%%%%%%%%%%%%%%%%%%%%%%%%%%%%%%%%%%%%%%%%%%%%%%%%%%%%%%%%%%%%%%%%%%%%%%
%
%               END OF REVISION
%
%%%%%%%%%%%%%%%%%%%%%%%%%%%%%%%%%%%%%%%%%%%%%%%%%%%%%%%%%%%%%%%%%%%%%%%%%%%%%%%%%%
%-----------------------------------------------------------------------------------------------------------
%\section*{References}
%-----------------------------------------------------------------------------------------------------------
\bibliographystyle{unsrt}
\bibliography{biblio}

\end{document}